\documentstyle[12pt,epsfig]{article}
                                                                                
\renewcommand{\thepage}{}

\begin{document}                                                                
\date{}
                                                                                
\title{                                                                         
{\vspace{-3cm} \normalsize                                                      
\hfill \parbox{50mm}{DESY 95-206    \\
                     CERN-TH/95-315 \\
                     ITP-Budapest Rep.~No.~514}}\\[25mm]
Numerical tests of the electroweak phase transition and    \\
thermodynamics of the electroweak plasma                   \\[8mm]}
\author{
F. Csikor  \\
Institute for Theoretical Physics, E\"otv\"os University,  \\
Budapest, Hungary                           \\[4mm]
{Z. Fodor}\thanks{On leave from
Institute for Theoretical Physics,              
E\"otv\"os University, Budapest, Hungary.}  \\
Theory Division, CERN, 
CH-1211 Geneva 23, Switzerland              \\[4mm]
J. Hein, A. Jaster, I. Montvay              \\
Deutsches Elektronen-Synchrotron DESY,      \\
Notkestr.\,85, D-22603 Hamburg, Germany}                                        
                                                                                
\newcommand{\be}{\begin{equation}}                                              
\newcommand{\ee}{\end{equation}}                                                
\newcommand{\half}{\frac{1}{2}}                                                 
\newcommand{\rar}{\rightarrow}                                                  
\newcommand{\lar}{\leftarrow}                                                   
                                                                                
\maketitle                                                                      
                                                                                
\begin{abstract} \normalsize                                                    
The finite temperature phase transition in the SU(2) Higgs model
at a Higgs boson mass $M_H \simeq 34$ GeV is studied in numerical
simulations on four-dimensional lattices with time-like extensions
up to $L_t=5$.
The effects of the finite volume and finite lattice spacing on masses
and couplings are studied in detail.
The errors due to uncertainties in the critical hopping parameter
are estimated.
The thermodynamics of the electroweak plasma near the phase transition
is investigated by determining the relation between energy density
and pressure.
\end{abstract}       
%
%
%
\newpage
\renewcommand{\thepage}{\arabic{page}}
\section{Introduction}\label{sec1}
 The electroweak phase transition \cite{KIRLIN} plays an important
 r\^ole in the baryon asymmetry of the Universe \cite{KURUSH}.
 Since near the phase transition and in the phase with restored
 symmetry infrared singularities render the perturbation theory
 uncertain, non-perturbative numerical simulations may be very useful
 in providing numerical control of the perturbative resummation
 techniques \cite{ARNESP,BFHW,FOHE,BUFOHE}.

 The present paper is a continuation of our previous work on this
 subject \cite{PHYSLETT,NUCLPHYS,CSFOHEHE}.
 Our main goal is to test different sources of systematic errors
 in four-dimensional simulations.
 The value of the Higgs boson mass $M_H \simeq 34\; GeV$ is chosen
 such that comparisons of our four-dimensional results with
 three-dimensional work in reduced models
 \cite{FAKARUSH,KANEPA,ILKRPESCH,JAPAPE} become possible.
 The dimensional reductional technique can be used to test classes
 of electroweak phase transition models; however, the quality 
 of the approximation should be controlled.
 The hope is that for Higgs boson masses between $20\; GeV$ and
 $80\; GeV$ the combined outcome of resummed perturbation 
 theory and of unreduced and reduced numerical simulations will
 be a complete understanding of the thermodynamics of the electroweak
 phase transition.
 (For a recent review of the subject, see e.~g.~\cite{JANSEN}.)
 In fact, in the present paper we also do a first step towards the
 determination of the thermodynamical equations of state in the
 electroweak plasma, by investigating the relation between energy
 density and pressure on both sides of the phase transition
 (for $1/4 \leq T/T_c \leq 2$).

 The notations in the present paper are the same as in our previous
 works \cite{PHYSLETT,NUCLPHYS,CSFOHEHE}.
 For the reader's convenience we repeat here the lattice action:
$$
S[U,\varphi] = \beta \sum_{pl}
\left( 1 - \frac{1}{2} {\rm Tr\,} U_{pl} \right)
$$
\be \label{eq01}
+ \sum_x \left\{ \half{\rm Tr\,}(\varphi_x^+\varphi_x) +
\lambda \left[ \half{\rm Tr\,}(\varphi_x^+\varphi_x) - 1 \right]^2 -
\kappa\sum_{\mu=1}^4
{\rm Tr\,}(\varphi^+_{x+\hat{\mu}}U_{x\mu}\varphi_x)
\right\} \ .
\ee
 Here $U_{x\mu}$ denotes the SU(2) gauge link variable, $U_{pl}$
 is the product of four $U$'s around a plaquette, and
 $\varphi_x$ is a complex $2 \otimes 2$ matrix in isospin space
 describing the Higgs scalar field.
 The bare parameters in the action are $\beta \equiv 4/g^2$ for
 the gauge coupling, $\lambda$ for the scalar quartic coupling and
 $\kappa$ for the scalar hopping parameter related to the bare
 mass square $\mu_0^2$ by $\mu_0^2 = (1-2\lambda)\kappa^{-1} - 8$.
 In what follows we set the lattice spacing to 1 ($a=1$),
 therefore all the masses and correlation lengths, etc., will always be
 given in lattice units, unless otherwise stated.

 The numerical simulations were performed on the APE-Quadrics
 parallel computers at DESY-Zeuthen.
 In order to obtain the desired value of the Higgs boson mass
 ($M_H \simeq 34$ GeV), we had to choose for the bare quartic
 coupling $\lambda \simeq 0.0003$.
 The other two bare couplings are in the same range as for the
 previously studied cases: $\lambda \simeq 0.0001,\; 0.0005$
 (corresponding to $M_H \simeq 18,\; 50$ GeV, respectively).
 The lattice extents in time (i.~e.\ inverse temperature) direction
 were in the range $2 \leq L_t \leq 5$.
 For the determination of the thermodynamical equation of state
 we used $L_t=2$ and 4.
 These choices of $L_t$ allowed us to estimate the magnitude of
 the lattice discretization errors, too.

 The simulation algorithms have been described in detail in our
 previous publications.
 An important ingredient for reducing the autocorrelations of the
 generated field configurations is the overrelaxation algorithm for
 the Higgs field \cite{FODJAN}.
 The numerical tests in our parameter range showed that the best
 variant is the one proposed in \cite{BUNK}.
 For instance, comparing the methods of \cite{FODJAN} and \cite{BUNK}
 on a $12^3 \cdot 32$ lattice with $\beta=8.0$,
 $\kappa=0.1287$ and $\lambda=0.0003$ the method of \cite{BUNK}
 reduced the integrated autocorrelation time for the length of the
 Higgs field from 7.4 sweeps to 0.9 sweeps.
 The speed-up factor in CPU-time is about 14, because the algorithm
 in \cite{BUNK} is simpler.
 
 The plan of this paper is as follows:
 First, in the next section, the finite volume effects on masses and
 couplings are discussed.
 The errors of the critical hopping parameter and their effects on
 the determination of the latent heat are investigated in
 section~\ref{sec3}, whereas section~\ref{sec4} is devoted to the
 thermodynamics of the electroweak plasma.
 In section~\ref{sec5} our results are compared to two-loop resummed
 perturbation theory.
 Finally, the last section contains the discussion of the results and
 future plans.

                                                                                
\section{Finite volume and finite lattice spacing effects}\label{sec2}
 In this section we discuss the finite volume and finite lattice
 spacing effects on the zero temperature renormalized quantities,
 i.~e.\ the masses and the gauge coupling.

  
\subsection{Zero temperature masses}                  \label{subsec21}
 As in refs. \cite{PHYSLETT,NUCLPHYS}, the physical Higgs mass $M_H$
 was extracted from correlators of the following quantities:
\begin{equation} \label{eq02}
R_x \equiv \frac{1}{2} {\rm Tr}(\varphi_x^+ \varphi_x) 
\equiv \rho_x^2 \,,
\end{equation}
 and, using $\varphi_x \equiv\rho_x \alpha_x$ with
 $\alpha_x \in {\rm SU(2)}$,
\begin{equation} \label{eq03}
L_{\alpha , x \mu} \equiv \frac{1}{2} 
{\rm Tr}(\alpha_{x+\hat{\mu}}^+ U_{x\mu}\,
 \alpha_x) \, ,
\hspace{3em} 
L_{\varphi , x \mu} \equiv \frac{1}{2} 
{\rm Tr}(\varphi_{x+\hat{\mu}}^+ U_{x\mu}\, 
\varphi_x) \, .
\end{equation}
 The W-boson mass $M_W$ was obtained from the correlator of the
 composite link fields 
\begin{equation} \label{eq04}
W_x  \equiv \sum_{r,k=1}^3 \frac{1}{2} {\rm Tr}
(\tau_r \alpha_{x+\hat{k}}^+ U_{xk}\, \alpha_x ) \, .
\end{equation}

 Simulations to determine the zero temperature masses were performed
 on several lattices of different space extensions at the phase
 transition point of the $L_t=2$ lattice.
 For $L_t=3,4,5$ the lattice volumes were chosen large enough, to
 be close to the infinite volume limit.
 The simulation parameters are collected in table~\ref{tab01}.
 Besides the simulation points close to the critical hopping parameter
 there are also points with shifted $\kappa$, in order to control
 the effects of uncertainties in the critical hopping parameter.
\begin{table*}[tb]
\begin{center}
\parbox{15cm}{\caption{\label{tab01}\it
 The parameter values of numerical simulations for the determination of
 zero temperature masses and renormalized gauge couplings.
 The indices for the four sets of simulation points {\rm m2, m3, m4}
 and {\rm m5} correspond to $L_t=2,3,4$ and $5$, respectively.
 The numbers in brackets refer to the linear size of the lattice and to
 the last two digits of the hopping parameter.}}
\end{center}
\begin{center}
\begin{tabular}{|c||c|c|l|l||r|c|}
\hline
index & lattice & $\beta$ & \multicolumn{1}{c|}{$\kappa$} &
\multicolumn{1}{c||}{$\lambda$} & sweeps & subsamples \\
\hline\hline
m2[6] & $6^3 \cdot  32$&8.00 & 0.12865 & 0.0003 & 384000 & 64 \\
m2[8] & $8^3 \cdot  32$&8.00 & 0.12865 & 0.0003 & 288000 & 16 \\
m2[12/65] & $12^3 \cdot  32$&8.00 & 0.12865 & 0.0003 & 320000 & 80 \\
m2[16] & $16^3 \cdot  32$&8.00 & 0.12865 & 0.0003 & 192000 & 96 \\
m2[24] & $24^3 \cdot  64$&8.00 & 0.12865 & 0.0003 & 38000 & 20 \\
m2[12/70] & $12^3 \cdot  32$&8.00 & 0.12870 & 0.0003 & 448000 & 32 \\
\hline
m3[24/35] & $24^3 \cdot  32$&8.15 & 0.128355 & 0.00031 & 32000 & 80 \\
m3[24/40] & $24^3 \cdot  32$&8.15 & 0.128405 & 0.00031 & 11200 & 28 \\
\hline
m4[24/35] & $24^3 \cdot  64$&8.25 & 0.128235 & 0.000315 &168000 & 28\\
m4[24/85] & $24^3 \cdot  64$&8.25 & 0.128285 & 0.000315 & 14400 & 36\\
\hline
m5[32/17] & $32^3 \cdot  64$&8.33 & 0.128170 & 0.000319 & 37740 & 74\\
m5[32/21] & $32^3 \cdot  64$&8.33 & 0.128210 & 0.000319 & 15300 & 30\\
\hline
\end{tabular} \end{center}
\end{table*}

 Masses were extracted from the correlators fitting to a cosh+constant
 function. 
 Simple uncorrelated least square fits 
 and correlated fits (eventually with eigenvalue smoothing)
 \cite{MICHAEL} were used.
 The use of the latter method (Michael--McKerrell method) is necessary,
 since the data are strongly correlated for different time distances. 

 For the reader's convenience -- following \cite{MICHAEL} -- we briefly
 define how to perform correlated and Michael--McKerrell fits.
 Let us denote the measured data of a correlator by 
 $x^{(n)}(t)$, with $t$ the time difference ($t=t_0+1\ldots t_0+D)$.
 We assume $N$ data samples (i.~e.\ $n=1\ldots N$), with averages
 $\overline{X}(t)$.
 Our aim is to fit our data to a given function $F(t,a)$,
 which depends on $P$
 parameters $a \equiv a_1,\ldots,a_P$. 
 To find the best fit parameters corresponds to minimizing with
 respect to $a$
\begin{equation} \label{eq05}
 \chi^2= \sum_{t,t'}
   (F(t,a)-\overline{X}(t)) M(t,t') (F(t',a)-\overline{X}(t')) \ ,
\end{equation}
 where
\begin{equation} \label{eq06}
\overline{X}(t)={1 \over N} \sum_{n=1}^N x^{(n)}(t) \ ,
\end{equation}
\begin{equation} \label{eq07}
M(t,t')=N C^{-1}(t,t') \ ,
\end{equation}
 with
\begin{equation} \label{eq08}
C(t,t')=    {1 \over N-1} \sum_{n=1}^N
(x^{(n)}(t)-\overline{X}(t))(x^{(n)}(t')-\overline{X}(t')) \ .
\end{equation}
 The uncorrelated fit corresponds to ignoring non-diagonal elements
 in the correlation matrix $C(t,t')$.

 The difficulty is that the elements of the inverse correlation matrix
 have a tendency to get large statistical errors.
 To deal with this problem, the Michael--McKerrell method proposes to
 change the $D$ eigenvalues $\lambda_i$  of the normalized correlation
 matrix
\begin{equation} \label{eq09}
 \tilde{C}(t,t') = { C(t,t') \over \sqrt{C(t,t) C(t',t')} }
\end{equation}
 as follows:
\begin{equation} \label{eq10}
  \lambda^{'}_{i} = K \ \hbox{max} ( \lambda_i, \lambda_{min}) \ ,
\end{equation}
 where
\begin{equation} \label{eq11}
\lambda_{min}= {1 \over D-E} \sum_{i=E+1}^{D} \lambda_i
 \ ,  \hspace{3em}
K^{-1}= { 1 \over D} \sum_{i=1}^{D}
\ \hbox{max} ( \lambda_i, \lambda_{min}) \ .
\end{equation}

 It is assumed that the eigenvalues $\lambda_i$ are arranged in
 decreasing order.
 The eigenvectors of $\tilde{C}(t,t')$ and thus of its inverse
 $\tilde{M}(t,t')$ are retained unchanged.
 The procedure removes any very small eigenvalues of
 $\tilde{C}(t,t')$ and replaces them with the average of the
 $D-E$ smallest eigenvalues while retaining the property that the
 trace of $\tilde{C}(t,t')$ is $D$.
 From the practical tests it is suggested that $E$ should be about
 $\sqrt N$.

 The advantage of the correlated (or Michael--McKerrell) method is
 twofold.
 First, the values of $\chi^2$ per degree of freedom obtained by the
 uncorrelated least square fits are  notoriously low, thus making
 impossible the choice of  the reasonable fit interval.
 Performing a correlated fit or its Michael--McKerrell extension,
 the $\chi^2$/d.o.f.\ is reasonable, i.~e.\ the best
 value is near unity.
 The second advantage is obtained when the statistics and the number
 of subsamples is low.
 In this case the correlation matrix $\tilde{C}(t,t')$ may have
 very low eigenvalues, which influences unreasonably the inverse of
 the correlation matrix used in modelling the distribution of the
 data.
 The proposal of \cite{MICHAEL} is to smear the smallest eigenvalues.
 It turns out that the method results in reasonable 
 $\chi^2$/d.o.f values even in such cases.
 Examples illustrating the above statements are given in
 tables~\ref{tab02} and \ref{tab03}.

 In order to perform the correlated (and Michael--McKerrell) fits,
 the data were subdivided into subsamples.
 The errors on the data in the subsamples are not used in these fits.
 The errors of the data in the total data sample (used in the 
 uncorrelated fits) were  determined from the 
 statistical fluctuations of subsample data or the errors of the 
 subsample's data.
 The statistical errors of masses obtained in the fits were
 determined by jackknife analyses.
\begin{table*}[tb]
\begin{center}
\parbox{15cm}{\caption{\label{tab02}\it
 Comparison of the uncorrelated fit (index I) and correlated fit 
 (index II) for the $16^3\cdot 32$ lattice ($\rho^2_x $ correlator), 
 with data divided into 96 subsamples. 
 The 0--16 interval is obviously excluded by the correlated fit, while
 superficially the uncorrelated fit seems to be reasonable.}}
\end{center}
\begin{center}
\begin{tabular}{|c|c|c|c|c|}
\hline
interval&Higgs mass I & $\chi^2/d.o.f.$ I&Higgs mass II &
 $\chi^2/d.o.f.$ II \\
\hline\hline
2--8 & 0.24027 &1.072$\cdot 10^{-3}$ &0.24207 & 1.08 \\
\hline
2--10 & 0.23989 &1.494$\cdot 10^{-3}$  &0.24277 & 1.08 \\
\hline
2--12 & 0.24073 &5.71$\cdot 10^{-3}$  & 0.24283 & 1.01 \\
\hline
2--14 & 0.24265 &2.48$\cdot 10^{-3}$  & 0.24272 & 0.965 \\
\hline
2--16 & 0.24369 &5.92$\cdot 10^{-3}$  & 0.24327  & 0.992 \\
\hline
0--16 & 0.24601 &0.209  & 0.26242 & 59.13 \\
\hline
\end{tabular} \end{center}
\end{table*}
%
\begin{table*}[tb]
\begin{center}
\parbox{15cm}{\caption{\label{tab03}\it
 Comparison of the correlated fits with different numbers of smeared 
 eigenvalues of the correlation matrix for
 the $24^3\cdot 64$ lattice ($\rho^2 $ correlator), with data divided into
 20 subsamples. 
 The fit interval is 2--26, so there are 6 zero eigenvalues. 
 Inversion of the correlation matrix is impossible, smearing of the 
 lowest eigenvalues is necessary. 
 The last row (5 exact eigenvalues retained) gives quite a reasonable
 fit.}}
\end{center}
\begin{center}
\begin{tabular}{|c|c|c|c|}
\hline
interval& smeared eigenvalues & Higgs mass & $\chi^2/d.o.f.$ \\
\hline\hline
2--26 & 7 & 0.24573 & 266.9 \\
\hline
2--26 & 8 & 0.24609 & 34.79 \\
\hline
2--26 & 10& 0.24571 & 6.25  \\
\hline
2--26 & 12& 0.24518 & 3.88  \\
\hline
2--26 & 14& 0.24727 & 2.88  \\
\hline
2--26 & 16& 0.24643 & 1.79  \\
\hline
2--26 & 18& 0.24570 & 1.87  \\
\hline
2--26 & 20& 0.24361 & 1.75  \\
\hline
\end{tabular} \end{center}
\end{table*}

 The actual procedure of extracting the mass parameters is the
 following. 
 First one determines the reasonable intervals for fitting the data. 
 The guideline is to choose as large an interval as possible with
 reasonable $\chi^2$/d.o.f.\ value.
 For this purpose correlated fits with eigenvalue smearing were
 used.
 The {\em best fit} value of the masses was taken to be the number
 given by the uncorrelated fit.
 In case of the Higgs mass, where three different correlators were
 measured, the final value of the Higgs mass was obtained
 by averaging the individual fit results.
 The errors on the masses were determined by jackknife analysis.
 The masses obtained by the correlated fits with eigenvalue smearing
 are in all cases well within the error bars of the uncorrelated fits.
 Thus the uncertainty of the choice of the best fit value is not too
 important.
 An additional uncertainty is caused by the choice of the fit
 interval.
 Even though it seems reasonable to include as many points as allowed
 by the $\chi^2$/d.o.f., this is not compulsory.
 One may e.~g.\ take the intervals with smallest $\chi^2$/d.o.f.'s. 
 Again the additional uncertainty  caused by such a choice of the
 interval is not really important.

 Our results are summarized in table~\ref{tab04}.
 The first 5 rows there refer to the same physical situation, only the
 lattice sizes are different.
 Thus the data are suitable to demonstrate finite size effects.
 One observes that the masses and the Higgs to W-boson mass ratio
 ($R_{HW}$) are different for the $6^3\cdot 32$ lattice as compared to
 the larger ones.
 The larger lattices yield, within errors, equal masses and $R_{HW}$,
 thus finite size effects are already small for these lattices.
 (In case of the $24^3\cdot 64$ lattice, the errors are much larger 
 than for the other lattices.)

 Comparing the points with slightly shifted $\kappa$ but otherwise
 same parameters one observes that the masses are changed; however
 $R_{HW}$ remains unchanged within errors.
%
\begin{table*}[tb]
\begin{center}
\parbox{15cm}{\caption{\label{tab04}\it
 Final results on the Higgs and W masses.
 The lattices are in the same order as above in table
 \protect\ref{tab01} for simulation parameters and are identified by
 the indices defined there.
 Higgs and W masses are first given in lattice units.
 $R_{HW}=M_H /M_W $.}}
\end{center}
\begin{center}
\begin{tabular}{|c||c|l||c|l||l|c|}
\hline
index &interval& \multicolumn{1}{c||}{$M_H$}&interval &
\multicolumn{1}{c||}{$M_W$} & \multicolumn{1}{c|}{$R_{HW}$}& $M_H$ (GeV) \\
\hline\hline
m2[6]       &2 -- 16 &0.2336(13)&2 -- 16  &0.5587(13)  &0.4181(33) &33.5 \\
m2[8]       &2 -- 16 &0.2480(12)&3 -- 11  &0.5573(28)  &0.4450(44) &35.7 \\
m2[12/65]   &2 -- 16 &0.2451(14)&2 -- 16  &0.5619(21)  &0.4362(41) &35.0 \\
m2[16]      &2 -- 16 &0.2436(21)&2 -- 16  &0.5644(21)  &0.4316(53) &34.6 \\
m2[24]      &2 -- 32 &0.2471(39)&2 -- 22  &0.5665(17)  &0.436(8)  &35.0 \\
m2[12/70]   &2 -- 16 &0.2585(9) &2 -- 16  &0.5888(13)  &0.4390(25) &35.2 \\
\hline
m3[24/35]   &2 -- 32& 0.1587(14)&2 -- 32  &0.3712(31)  &0.428(7)  &34.2 \\
m3[24/40]   &2 -- 32& 0.1796(40)&2 -- 32  &0.414(8)  &0.433(18) &34.7 \\
\hline
m4[24/35]   &2 -- 32 &0.1198(15)&4 -- 32  &0.2765(23)  &0.433(9) &34.8 \\
m4[24/85]   &2 -- 32 &0.1428(25)&2 -- 32  &0.3427(42)  &0.417(12)  &33.4 \\
\hline
m5[32/17] &2 -- 32 & 0.0907(15) &4 -- 32 &0.221(5) & 0.411(16) & 32.9\\
m5[32/21] &2 -- 32 &0.1205(20)  &2 -- 32  &0.284(5)    &0.425(15) & 34.0\\
\hline
\end{tabular} \end{center}
\end{table*}

\begin{figure}
\epsfig{file=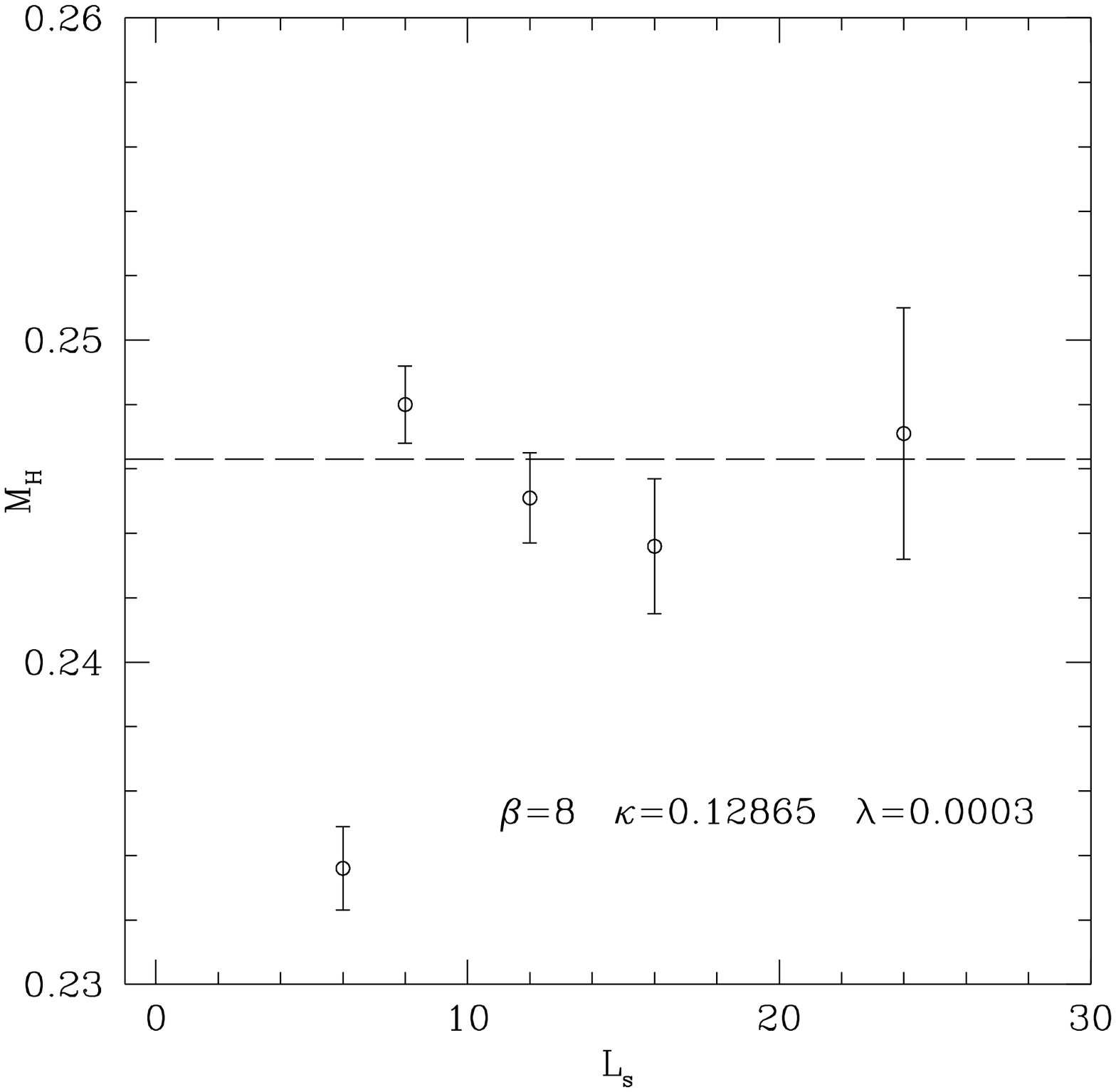,width=16.0cm,height=14.0cm,angle=0}
\begin{center}
\parbox{15cm}{\caption{ \label{Mh}
 The Higgs boson mass on different volumes.
 The horizontal line corresponds to $M_H=0.2463$.
}}
\end{center}
\end{figure}

\begin{figure}
\epsfig{file=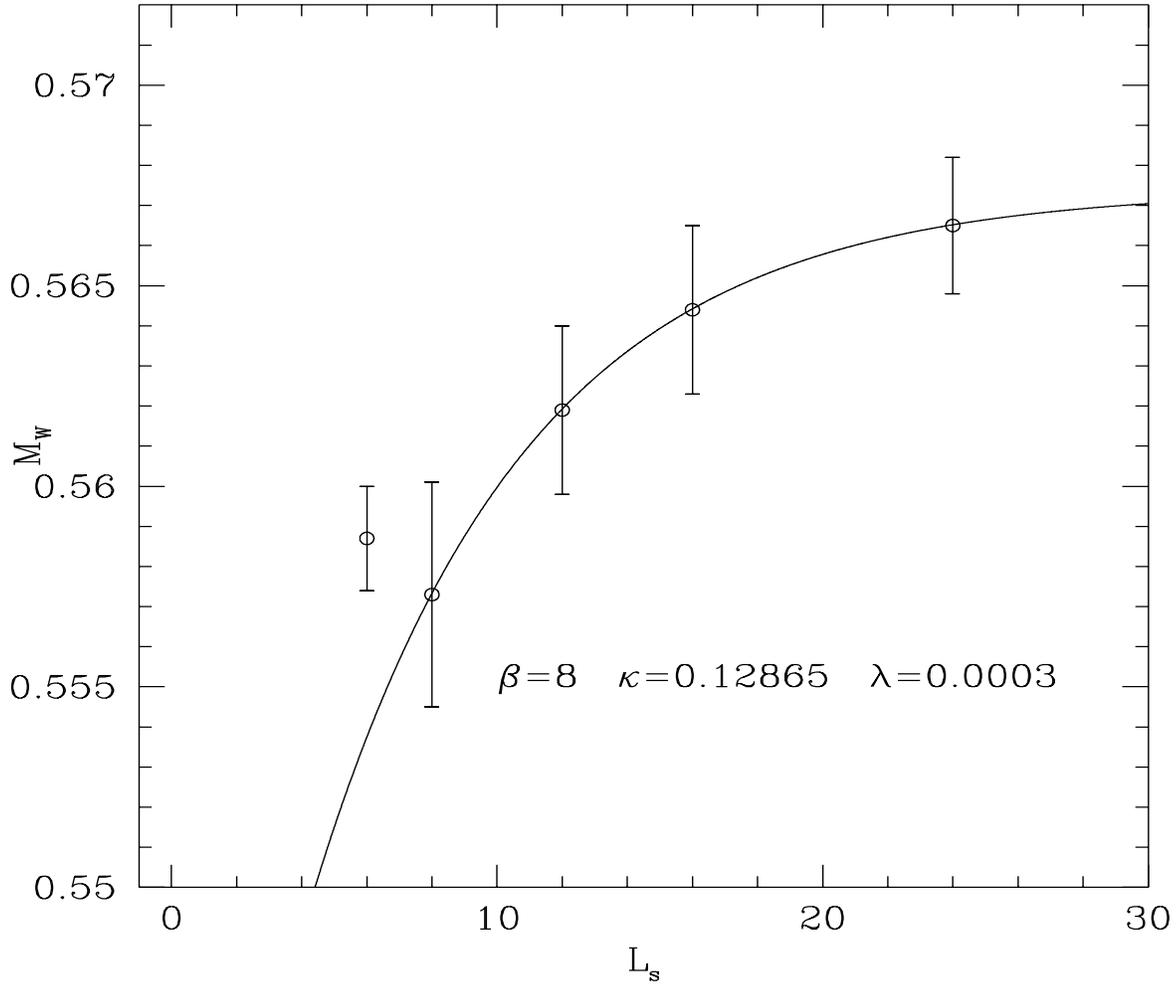,width=16.0cm,height=14.0cm,angle=0}
\begin{center}
\parbox{15cm}{\caption{ \label{Mw1}
 The W-boson mass on different volumes.
 The curve shows the three-parameter fit
 $M_W=(1-0.06\cdot \exp(-0.1523\cdot L_s))\cdot 0.5674$.
}}
\end{center}
\end{figure}

 For the parametrization of the volume dependence of masses one
 can apply a formula suggested by the large volume asymptotic behaviour
 in scalar field theory \cite{LUMOWE}.
 Here we only want to illustrate the qualitative behaviour, therefore
 we omit possible power corrections, and for spatial lattice extent
 $L_s$ assume the behaviour
\be \label{eq12}
M(L_s) \simeq M(\infty) [1-c\exp(-L_s m)] \ .                 
\ee
 This is a three-parameter form, with the infinite volume mass
 $M(\infty)$, the constant $c$ and the exchange mass parameter $m$.
 Since the number of points in $L_s$ is not much larger than the
 number of parameters, the only question is the qualitative
 description with a reasonable set of parameters.
 In fact, within our range of $L_s$ and statistical errors the
 Higgs boson mass is constant.
 This is illustrated by figure~\ref{Mh}.
 For the W-boson mass a reasonable fit of the form (\ref{eq12}) can be
 obtained (see figure~\ref{Mw1}).
 In this case the value of the exchange mass $m$ comes out, as
 expected, not much different from the Higgs mass $M_H$.

\subsection{Renormalized gauge coupling}         \label{subsec22}
 The renormalized gauge coupling was determined in the usual way from
 the static potential \cite{NUCLPHYS}.
 The potential as a function of the distance $R$ was fitted by
\be \label{eq13}
V(R)=-\frac{A}{R} e^{-M R}+C+D\, G(M,R,L_s) \ ,               
\ee
 where the last term with $G(M,R,L_s)$ corrects for lattice artefacts.
 The value of the potential at $R$ was obtained from the rectangular
 Wilson loops by fitting the time dependence with three exponentials.
 The high statistics for the Wilson loops allowed a stable fit with
 good $\chi^2$ if the smallest time distances, between one at
 $L_t=3$ and three at $L_t=5$, were omitted.
 The potential was then fitted by the form in (\ref{eq13}).
 For $L_t=2$ every $R$-value was used, but for $3 \leq L_t \leq 5$
 the first point with $R=1$ contributed too much to $\chi^2$.
 The discrepancy was increasing for increasing $L_t$.
 Fitting only $R \geq 2$ the value of the fit curve at $R=1$ deviated
 from the measured one by about hundred standard deviations.
 Therefore the formula (\ref{eq13}) is clearly not valid for
 $L_t \geq 3$ and $R=1$.
 The tree-level perturbative correction for lattice artefacts
 $G(M,R,L_s)$ is not good enough for our high precision data.
 Omitting also $R=2$ and fitting only for $R \geq 3$ gave already
 compatible results with $R \geq 2$.
 Hence the fits for $3 \leq L_t \leq 5$ were done with $R \geq 2$.

\begin{figure}
\epsfig{file=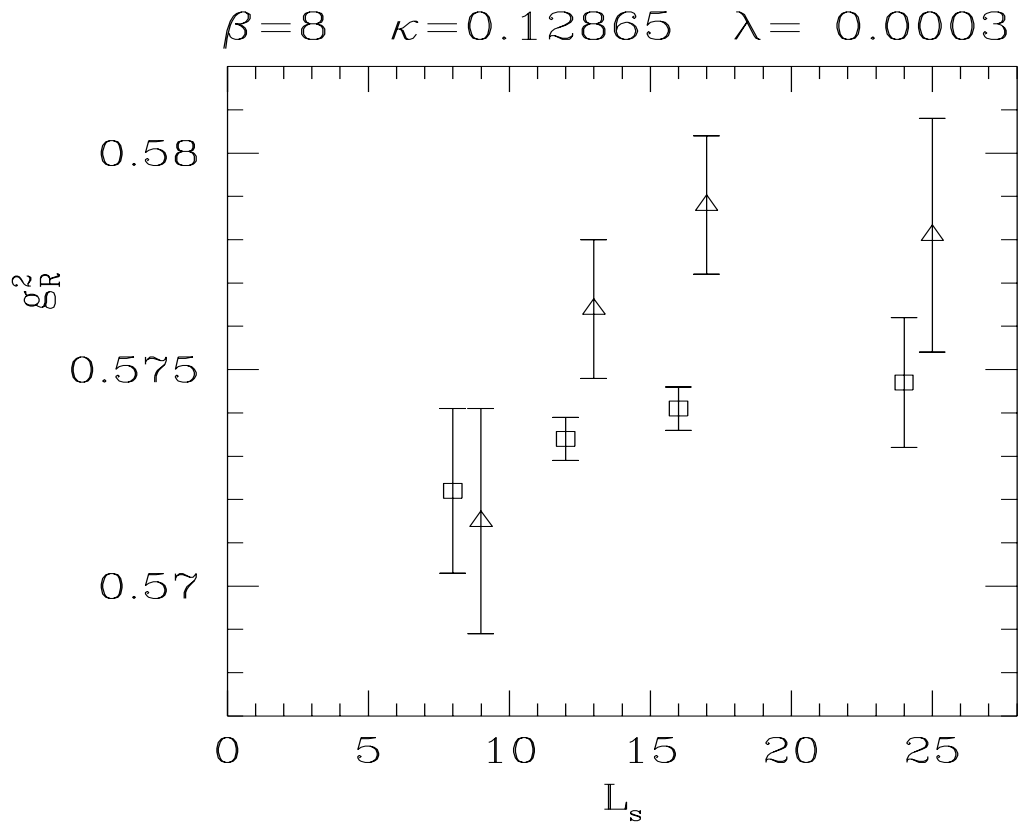,width=15.0cm,height=12.5cm,
        bbllx=0pt,bblly=120pt,bburx=590pt,bbury=730pt,angle=0}
\begin{center}
\parbox{15cm}{\caption{ \label{FigFinSiz1}
 Finite size effects for the renormalized gauge coupling.
 The squares represent the values obtained by the global definition.
 The values for the local definition are the triangles, which are
 slightly displaced for better  visualization.
}}
\end{center}
\end{figure}

\begin{figure}
\epsfig{file=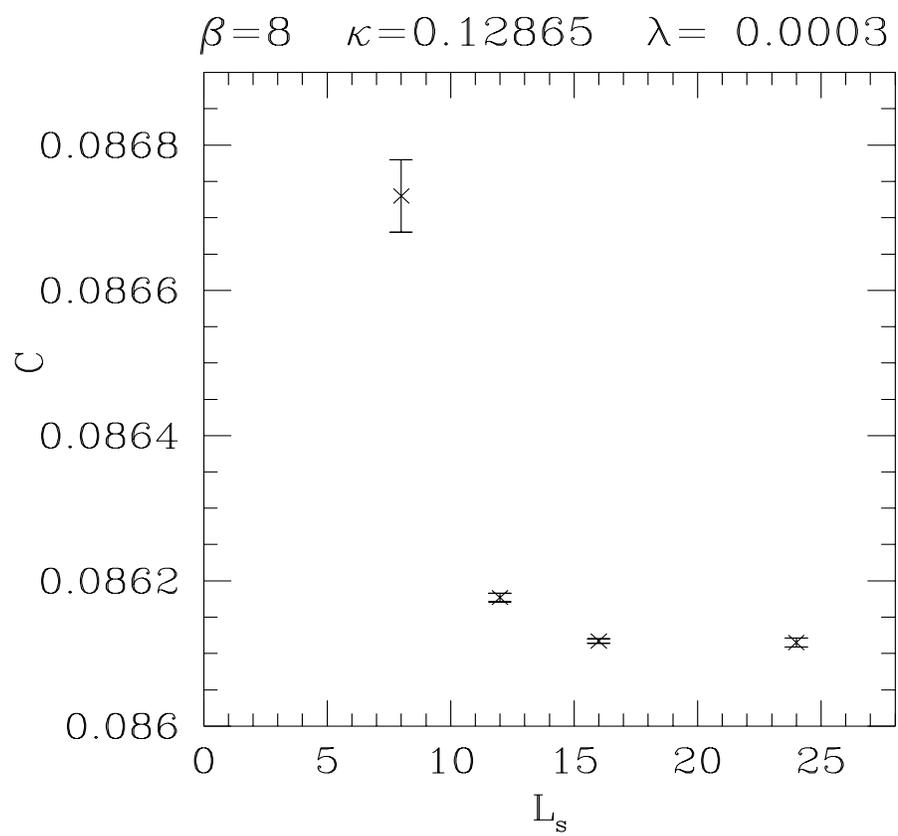,width=17.0cm,height=12.0cm,
        bbllx=0pt,bblly=0pt,bburx=420pt,bbury=320pt,angle=0}
\begin{center}
\parbox{15cm}{\caption{ \label{FigFinSiz2}
 Finite size effects for the constant $C$ in the static potential.
}}
\end{center}
\end{figure}

\begin{figure}
\epsfig{file=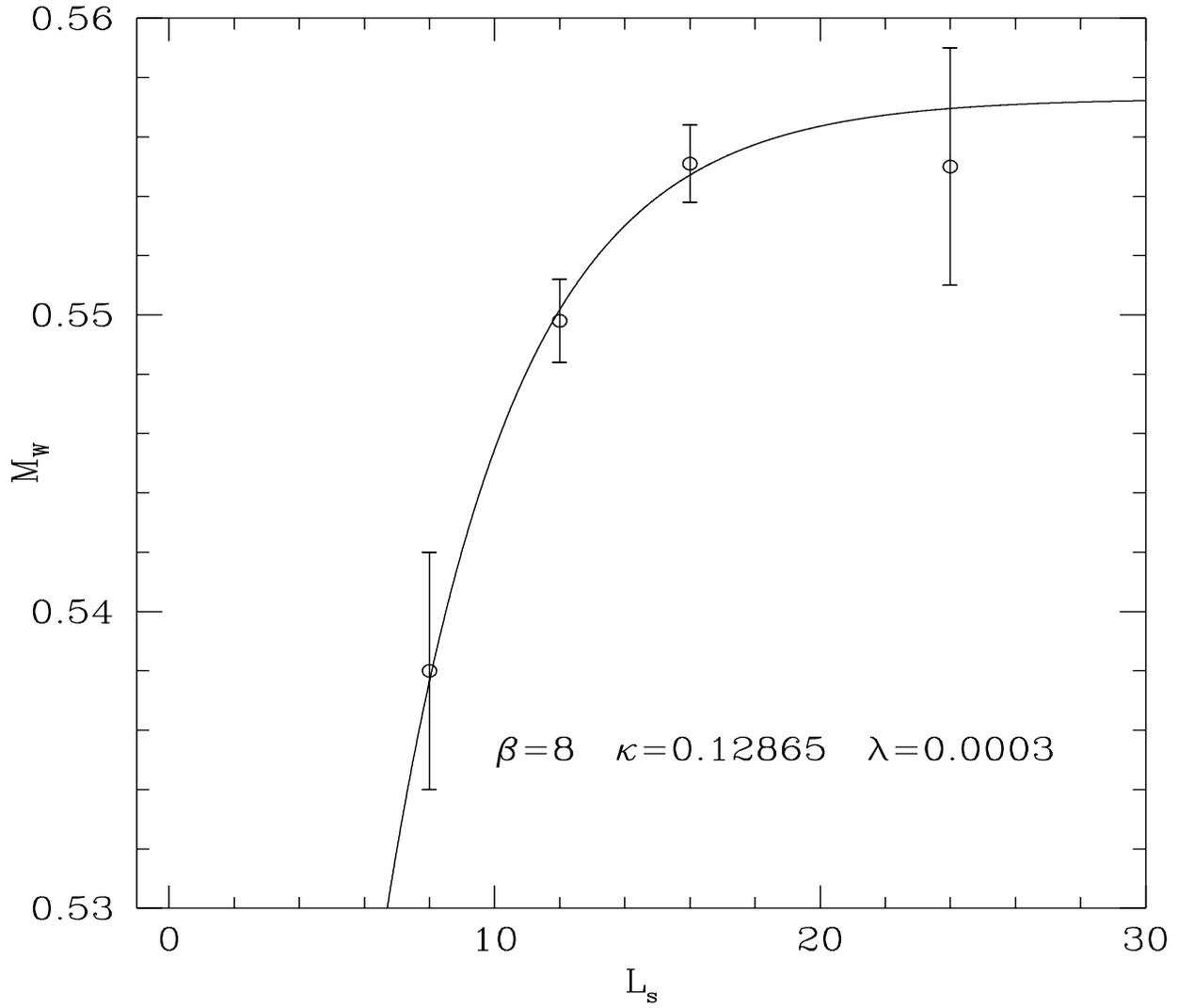,width=17.0cm,height=15.0cm,angle=0}
\begin{center}
\parbox{15cm}{\caption{ \label{Mw2}
 The W-boson mass on different volumes as obtained from the static
 potential.
 The curve shown is the three-parameter fit
 $M=(1-0.2678\cdot \exp(-0.2535\cdot L_s))\cdot 0.5573$.
}}
\end{center}
\end{figure}

 All results are collected in table~\ref{tab05}, with the point
 indices given in table~\ref{tab01}.
 The point m2[6] was omitted, because at least $L_s=8$ is required
 to fit four parameters.
 Obviously, finite size effects for the renormalized gauge coupling
 and the screening mass are small.
 The reason is that the values of these parameters are determined
 from the behaviour of the potential at small distances.
 Only the constant $C$, which is given by the  asymptotic dependence,
 shows larger finite size effects.
 The global ($g_R^2$) and local ($g_R^2(M^{-1})$) gauge couplings 
 (defined e.~g.\ in \cite{NUCLPHYS}) and the constant $C$ as a function
 of the space extention $L_s$ are shown in figs.~\ref{FigFinSiz1} and
 \ref{FigFinSiz2}.
 The volume dependence of the W-boson mass parameter $M$ obtained from
 the static potential is comparable with that of the pole mass $M_W$.
 This is shown in figure~\ref{Mw2}, where a fit of the same form as
 in eq.~(\ref{eq12}) is plotted.
 The values of the three fit parameters are roughly the same as
 those for $M_W$ in figure~\ref{Mw1}.
%
\begin{table}[tb]
\begin{center}
\parbox{15cm}{\caption{\label{tab05}\it
 Summary of the fit parameters for the static potential  and 
 the renormalized gauge coupling.}}
\end{center}
\begin{center}
\begin{tabular}{|c||l|l|l|l|l||l|}
\hline
index   & \multicolumn{1}{c|}{$A$} & \multicolumn{1}{c|}{$M$}  &
          \multicolumn{1}{c|}{$D$} & \multicolumn{1}{c|}{$C$}  &
          \multicolumn{1}{c||}{$g_R^2 \equiv {16 \over 3}\pi A$} &
          \multicolumn{1}{c|}{$g_R^2(M^{-1})$}                 \\
\hline\hline
m2[8]   & 0.03415(11)  &  0.538(4)  &                                 
  0.0336(7)   &  0.08673(5)  &  0.5722(19) & 0.571(3) \\                
m2[12/65]   & 0.03422(3)  &  0.5498(14)  &                                 
  0.0356(5)   &  0.086177(6)  &  0.5734(5) & 0.5764(16) \\                
m2[16]   & 0.03427(3)  &  0.5551(13)  &                                 
  0.0366(6)   &  0.086117(3)  &  0.5741(5) & 0.5788(16) \\                
m2[24]   & 0.03430(9)  &  0.555(4)  &                                 
  0.0361(18)   &  0.086115(6)  &  0.5747(15) & 0.578(3) \\                
m2[12/70] & 0.03420(3)  &   0.5815(14) &   
  0.0360(5)  &   0.085050(5)  & 0.5731(5) & 0.5764(13)  \\
\hline
m3[24/35] & 0.0367(11)  &   0.378(7) & 
  0.026(4)  &   0.091716(12)  & 0.615(18) & 0.591(7)  \\
m3[24/40] &0.0354(23)   &   0.413(13) &
  0.031(9) &   0.090100(15)  & 0.602(38) & 0.583(10) \\ 
\hline
m4[24/35]       & 0.0353(6)  &  0.269(5) &
  0.0293(24)    &  0.094267(19)  & 0.592(10)  &  0.584(6)  \\
m4[24/85]       & 0.0352(10)  &  0.328(7) &
  0.030(4)    &  0.092065(21)  & 0.590(17)  &  0.580(8)  \\
\hline
m5[32/17]      &   0.0354(5)   & 0.2135(35)  &
  0.0275(22)   &    0.095416(20)  & 0.593(8) & 0.586(8)\\
m5[32/21]     &   0.0352(7)   & 0.270(5)    & 
  0.028(3)   &   0.093295(22)  & 0.590(12)   & 0.581(8) \\ \hline
\end{tabular} \end{center}
\end{table}

 In summary, the finite size effects on our zero temperature lattices
 are small.
 For instance, at the phase transition point of the $L_t=2$ lattices
 the values of the Higgs-boson and W-boson masses as well as those of
 the renormalized gauge coupling are on lattices with spatial
 extension $12^3$ equal within errors to the values on $16^3$ and
 $24^3$.
 Therefore, these latter lattice sizes can be considered, within our
 small statistical errors, to represent the infinite volume situation.
 This information obtained on $L_t=2$ lattices can be used, by
 scaling up $L_s$ with $L_t$, to choose the spatial lattice size of
 $L_t \geq 3$ lattices sufficiently large, in order to avoid finite
 volume effects.

 Another important question is the size of lattice artefacts.
 These are expected to be of the order ${\cal O}(1/L_t^2)$, hence
 decrease by a factor $4/25$ between $L_t=2$ and $L_t=5$.
 Our points with increasing $L_t$ were chosen on a line of constant
 physics (LCP) using the one-loop $\beta$-functions \cite{NUCLPHYS}.
 Along such lines $R_{HW}$ and $g_R^2(M^{-1})$ should be constant.
 An investigation of tables \ref{tab04} and \ref{tab05} shows that,
 within errors, this is indeed the case: all values at the critical
 hopping parameter on large lattices agree with
 $R_{HW}=0.422(11)$ and $g_R^2(M^{-1})=0.585(10)$.
 At the same time the masses scale properly by $1/L_t$.
 In addition, the physical fit parameters in the potential, namely
 $g_R^2(M^{-1})$ and $M/M_W$ are also constant within errors,
 therefore also the physical shape of the potential scales.
 It is also true that the mass parameter $M$ obtained from the static
 potential is equal to the pole mass $M_W$ of the $W$ boson.
 In fact, the good scaling of the masses and static potential already
 at $L_t=2$ is surprising.
 Since our errors are at the level of 1-2\%, the magnitude of lattice
 artefacts in $R_{HW}$ and $g_R^2(M^{-1})$ on $L_t=5$ lattices is at 
 the level of a few parts per thousand.
  
\section{Critical hopping parameter and latent heat}    \label{sec3}
 An important step in numerical simulations is the determination of
 the critical hopping parameter $\kappa_c$ on lattices with high
 temperatures, where the first order phase transition between the
 Higgs phase and the symmetric phase (i.~e.\ phase with restored
 symmetry) occurs.
 On lattices with finite spatial volumes there is some uncertainty
 concerning the exact definition of this critical hopping parameter
 value.
 Usually, different possible definitions, as the equal area or equal
 height criterion in different order parameter distributions, give
 slightly different values.
 For growing volumes this uncertainty goes rapidly (in most cases
 exponentially) to zero, but in finite volumes this constitutes an
 inherent uncertainty.
 In addition, the statistical errors of the numerical simulations
 also increase the errors of $\kappa_c$.

 This error propagates to other physical quantities as, for instance,
 latent heat.
 The reason for this is that the uncertainty in the knowledge
 of the exact position of the transition point leads to additional
 uncertainties in the magnitude of the jump of order parameters like
 $\Delta P_{pl}$, $\Delta L_{\varphi,x\mu}$, etc.
 In order to control this kind of errors, in the present
 section we discuss how to determine the $\kappa$-dependence of
 the order parameters in a fairly wide $\kappa$-range from the results
 of a single run.
 In the determination of the latent heat, this procedure is used for
 a refined error analysis.

\subsection{Critical point and renormalization group trajectories}
                                                     \label{sec31}
 As discussed in refs.\ \cite{PHYSLETT,NUCLPHYS}, the knowledge of
 the critical hopping parameters can be directly used to 
 determine the renormalization group trajectories or {\em lines
 of constant physics} (LCP's).
 The derivatives of the bare parameters along LCP's also appear in the
 formula for the latent heat (see section~\ref{sec33}).

 For the flow of the bare couplings $\beta$ and $\lambda$ along the
 lines of constant physics we used the one-loop perturbative
 renormalization group equation as already described in
 refs.~\cite{PHYSLETT,NUCLPHYS}. 
 In order to determine the critical hopping parameters, two different
 approaches were applied. 
 For $L_t=2$ we used the more precise constrained simulation,
 for which $\kappa_c$ is defined by the flat distribution of an order
 parameter between the two peaks of the first order phase transition.
 The method and the result are published in ref.~\cite{CSFOHEHE}.
 For the sake of completeness, we just quote the result for
 $\kappa_c$.
 This method was not applicable in the case of the larger
 $L_t$-values, because it requires 64-bit numerical precision,
 which is missing on the APE-Quadrics computer.
 The CRAY-YMP at HLRZ J\"ulich has the necessary precision,
 but due to memory restrictions we were not able to perform the
 simulations on large volumes, corresponding to $L_t=4$.
 Therefore the two-coupling method described in
 refs.~\cite{POTREB,NUCLPHYS} was used for $L_t>2$.
 This method devides the lattice into two halves.
 In one half $\kappa_1<\kappa_c$, in the other one $\kappa_2>\kappa_c$.
 Demanding two bulk phases and two interfaces one receives upper and
 lower bounds for $\kappa_c$.
 This is rather robust and simple but, of course, in the 32-bit
 arithmetics of APE-Quadrics it gives less precision.
 The obtained values for the critical hopping parameter
 $\kappa_c$ are given in table~\ref{tab06}.
\begin{table}
\begin{center}
\parbox{15cm}{\caption{\label{tab06}\it
 Critical hopping parameters.}}
\end{center}
\begin{center}
\begin{tabular}{|r@{$\cdot$}c@{$\cdot$}l|r@{.}l|r@{.}l
||r@{.}l|c|}\hline
\multicolumn{3}{|c|}{lattice}& \multicolumn{2}{c|}{$\beta$}&
\multicolumn{2}{c||}{$\lambda$}&\multicolumn{2}{c|}{$\kappa_c$}&
method\\ \hline\hline
2 & $24^2$ & 256& 8&0  & 0&0003   & 0&1286565(7)& constrained\\ \hline
3 & $32^2$ & 512& 8&15 & 0&00031  & 0&128355(5) & 2-coupling\\ \hline
4 & $44^2$ & 512& 8&25 & 0&000315 & 0&128235(5) & 2-coupling\\ \hline
5 & $56^2$ & 560& 8&33 & 0&000319 & 0&128170(5) & 2-coupling\\ \hline
\end{tabular}
\end{center}
\end{table}

 At the phase transition points the temporal extension of the lattice
 is equal to the inverse transition temperature in lattice units:
 $L_t=1/T_c$.
 On a given LCP the ratio $M_W/T_c=M_W L_t$ is constant, therefore the
 change of the scale parameter $\tau\equiv-\log(M_W)$ between
 $L_t=L_t^{(1)}$ and $L_t=L_t^{(2)}$ is given by
 $\Delta\tau \equiv \tau^{(2)}-\tau^{(1)}=\log(L_t^{(2)}/L_t^{(1)})$.
 Since the lattice spacing is set to $a=1$, $\tau$ characterizes the
 lattice resolution.
 With increasing $\tau$ the continuum limit is approached.

 The derivative of the hopping parameter with respect to $\tau$ along
 the LCP's can be obtained from polynomial interpolations of the
 simulation data for $\kappa_c$ on lattices with different time
 extensions $L_t$.
 Here we consider lattices with $2 \leq L_t \leq 5$.
 In order to determine the statistical errors of the derivatives,
 the polynomial interpolations were repeated with 500 normally
 distributed random values around the measured mean values.
 An estimate of the systematic errors can be obtained by comparing
 2nd- and 3rd-order interpolations.
 The resulting values for $\partial\kappa/\partial\tau$ including
 both types of errors are:
\begin{equation}\label{eq14}
\left.\frac{\partial \kappa}{\partial\tau}\right|_{\rm m2}
     = -1.00(10)\cdot10^{-3},\qquad
\left.\frac{\partial \kappa}{\partial\tau}\right|_{\rm m3}
     = -0.53(4)\cdot10^{-3},\qquad
\left.\frac{\partial \kappa}{\partial\tau}\right|_{\rm m4}
     = -0.34(2)\cdot10^{-3} \ .
\end{equation}

\subsection{Uncertainties of the critical point and order parameters}
                                                        \label{sec32}
 As noted above, the uncertainties of the critical hopping parameters
 contribute to the errors of some other physical quantities.
 The variation of global quantities (e.~g.\ average link) leads to an
 important uncertainty in the determination of the latent heat.
 In principle, the dependence of these global quantities on $\kappa$
 can be determined in a single run by $\kappa$-reweighting
 \cite{FERREN}.
 This works well, if the $\kappa$-shifts are so small that the
 corresponding shifts of the measured quantities are smaller
 than their variances.
 However, in our case, due to the large lattice volumes, the allowed
 $\kappa$-shifts turned out to be too small.
 To come around this problem, we performed Taylor fits to the data
 obtained from reweighting and used the obtained Taylor series
 extrapolations to estimate the quantities at the required $\kappa$'s.

\begin{figure}
\epsfig{file=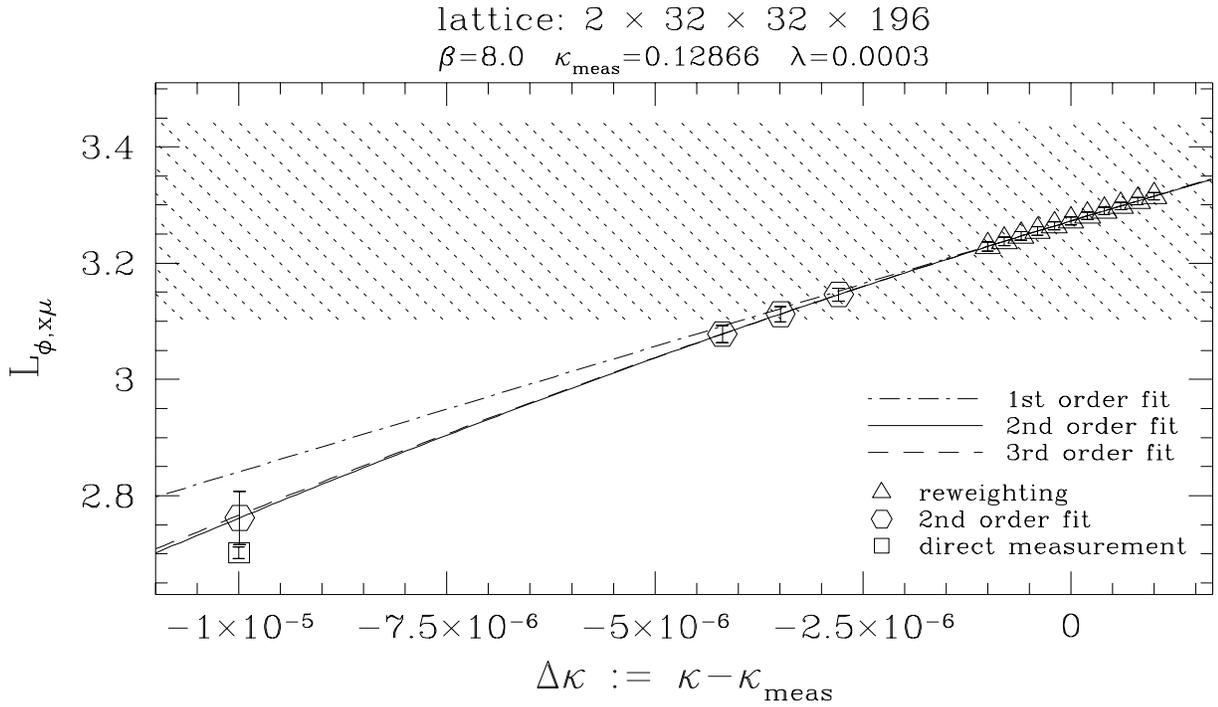,width=17.0cm,height=95mm,
        bbllx=77pt,bblly=404pt,bburx=566pt,bbury=686pt,angle=0}
\begin{center}
\parbox{15cm}{\caption{  \label{tprojfig}
 $L_{\varphi,x\mu}$ as a function of $\kappa$.
 The curves are fitted to the data points from the reweighting. 
 To give an impression of the errors some estimates from 2nd order
 are included. 
 The shaded area gives the variance of $L_{\varphi,x\mu}$ at
 $\kappa=0.12866$.
}}
\end{center}
\end{figure}

 We checked this procedure for the link variable by performing
 simulations with different $\kappa$'s on a
 $2\cdot 32\cdot 32\cdot 196$ lattice at $\beta=8.0$,
 $\lambda=0.0003$.
 The values of the hopping parameter were $\kappa=0.12865$ and
 $\kappa=0.12866$, respectively.
 Both of these points are in the metastable range.
 The lattice configurations were set into the Higgs phase.
 At $\kappa=0.12866$ we used reweighting to nearby $\kappa$ values.
 The result of the reweighting and the Taylor fits for the quantity
 $L_{\varphi,x\mu}$ is given in fig.~\ref{tprojfig}.
 The statistical errors of the Taylor coefficients and of the
 estimates at the required $\kappa$-values were determined by a
 bootstrap procedure \cite{GUPTA,CSFOHEHE} with 32 data blocks.

 Figure~\ref{tprojfig} shows that the first-order fit is insufficient
 to reproduce the result of the direct measurement at
 $\kappa=0.12865$.
 The second-order fit is in reasonably good agreement with the direct
 value, and the third-order fit gives only an insignificant change.
 This holds for the quantities $P_{pl}$ and $Q_x$ as well.
 Repeating this procedure in the symmetric phase, reasonable agreement
 within errors was found too.
 Therefore, also in other points the Taylor series was used up to
 the last significant coefficient.

\subsection{Latent heat}                                \label{sec33}
 As it has been already discussed in ref.~\cite{NUCLPHYS},
 the latent heat $\Delta\epsilon$ can be obtained from the
 discontinuity of $\delta \equiv \epsilon/3 - P$, where
 $\epsilon$ and $P$ denote energy density and pressure, respectively.
 This implies that $\Delta\epsilon$ can be determined from the jumps
 of some order parameters by
\begin{equation}\label{eq15}
\frac{\Delta \epsilon}{T_c^4} 
= L^4_t \left(8\frac{\partial \kappa}{\partial \tau}
             \langle \Delta L_{\varphi,x\mu}\rangle
  - \frac{\partial \lambda}{\partial \tau}
             \langle \Delta Q_{x}\rangle
  - 6\frac{\partial \beta}{\partial \tau}
             \langle \Delta P_{pl}\rangle \right) \ .
\end{equation}
 The link variable is defined in (\ref{eq03}), moreover 
\begin{equation}\label{eq16}
P_{pl}\equiv 1 - {\textstyle\frac{1}{2}}{\rm Tr}\, U_{pl} \ ,
\hspace{2em}
Q_x \equiv \left(\rho_x^2-1\right)^2 \ .
\end{equation}

 The results of the numerical simulations for the determination of
 the latent heat are given in table~\ref{tab07}.
 The order parameters in the vicinity of the transition point were
 obtained by the method described in the previous subsection.
 Note the sensitivity of the Higgs phase results on the small
 deviations in $\kappa$.
\begin{table}
\begin{center}
\parbox{15cm}{\caption{\label{tab07}\it
 Global average quantities for the determination of $\delta$ and
 latent heat.
 The first part refers to $T_c=\frac{1}{2}$, the second one to 
 $T_c=\frac{1}{4}$ (in lattice units).
 The bare parameters are $\beta=8.0, \lambda=0.0003$ and
 $\beta=8.15, \lambda=0.000315$, respectively. 
 At the phase transition there are two points corresponding to the
 two metastable phases.
 The statistical errors are quoted in parentheses.
 The last entry in every column gives the variation of the quantity
 in last digits when $\kappa$ is changed within its statistical
 error: $\kappa=0.1286565(\pm 7)$ and $\kappa=0.128235(\pm 5)$,
 respectively.
 The column ``fit'' specifies the order of the applied extrapolation.
}}
\end{center}\begin{center}
\begin{tabular}{|r@{$\cdot$}c@{$\cdot$}l||c|r@{.}l|r@{.}l|r@{.}l|}
\hline
\multicolumn{3}{|c||}{lattice} & fit 
& \multicolumn{2}{c|}{$P_{pl}$} 
& \multicolumn{2}{c|}{$L_{\varphi,x\mu}$}
& \multicolumn{2}{c|}{$Q_{x}$} \\ \hline \hline
32& $32^2$ & 32 & 1st order 
  &  0&0919035(8)$\mp61$&  9&1869(9)$\pm130$ & 132&124(21)$\pm300$ \\
\hline
 8&$ 32^2 $& 32 & 1st order
  & 0&0919075(7)$\mp61$ & 9&1757(8)$\pm 130$ & 131&898(18)$\pm 300$\\
\hline
6 &$32^2$& 96 & 1st order 
  &  0&0919204(9)$\mp61$& 9&1298(11)$\pm 130$& 130&888(25)$\pm 300$\\
\hline
5 &$32^2$& 96 & 1st order 
  & 0&0919557(10)$\mp 61$& 9&0521(11)$\pm 130$ & 
129&162(24)$\pm 300$ \\
\hline
4 &$32^2$& 96 & 1st order
  & 0&0920461(5)$\mp63$ &  8&8239(5)$\pm 130$  & 
124&103(10)$\pm 300$ \\ 
\hline
3 &$32^2$& 96 & 1st order
  & 0&0923641(5)$\mp70$ &  8&0267(6)$\pm 140$ & 
107&063(13)$\pm 300$ \\ 
\hline
2 &$32^2$& 96 & 2nd order
  &  0&094712(7)$\mp 19$ & 3&112(13)$\pm 34$ 
&  28&20(15)$\pm 39$ \\ 
\hline
2 &$32^2$& 96 & 2nd order
  & 0&0960210(6)$\mp3$ &  0&9199(4)$\pm 5$ &   8&074(3)$\pm 4$ \\ 
\hline \hline
48& $48^2$ & 48 & 1st order 
  &  0&0923167(3)$\mp 470$ &  2&5434(5)$\pm950$ &  
21&199(5)$\pm930$ \\ 
\hline
12& $64^2$ & 64 & 1st order
  &  0&0923226(4)$\mp 490$ &  2&5292(6)$\pm980$ &  
21&066(6)$\pm960$ \\ 
\hline
 8& $64^2$ & 64 & 1st order
  & 0&0923604(7)$\mp 530$ &  2&4571(10)$\pm 1000$ &
20 & 380(10)$\pm1000$ \\ 
\hline
 6 & $64^2$ & 192 & 1st order
  & 0&0924439(6)$\mp 500$ &  2&2809(9)$\pm1000$ & 
18&734(8)$\pm940$    \\ 
\hline
 5 & $64^2$ & 192 & 1st order
  & 0&0925683(6)$\mp 590$ &  2&0276(10)$\pm 1200$ & 
16&467(9)$\pm1000$   \\ 
\hline
 4 & $64^2$ & 192 & 2nd order
  & 0&0929636(11)$ {-1000 \atop +1300}$ & 1&
2440(23)$ {+2000\atop -2400}$ & 10 &227(17)$ {+1500\atop-1700}$ \\
\hline
 4 & $64^2$ & 192 & 2nd order
  & 0&0933086(3)$ {- 28 \atop +13}$  &  0&
5939(3)$ {+57\atop -32}$ & 5 & 955(2)$ {+34 \atop -19}$ \\ 
\hline
 3& $64^2$ & 192 & 2nd order
  &  0&0932657(4)${-8\atop +3}$ &  0&64536(7)$\pm110$ &  
6&2639(4)$\pm 69$     \\ 
\hline
 3& $92^2$ & 192 & 2nd order
  &  0&0932655(6)$\mp10 $ &  0&64522(14)$\pm100$ &  
6&2630(9)$\pm 63$     \\ 
\hline
  2& $64^2$ & 192 & 2nd order
  &  0&0930844(6)${-7\atop +1}$ &  0&83147(6)$\pm 75$ &  
7&4490(4)$\pm 52$     \\ 
\hline 
\end{tabular}
\end{center}
\end{table}

 The derivatives ${\partial \lambda}/{\partial \tau}$ and 
 ${\partial \beta}/{\partial \tau}$ in eq.~(\ref{eq15}) were taken
 from the one-loop perturbative renormalization group equations.
 As discussed in section~\ref{sec2}, the renormalized quantities show
 a good scaling between $L_t=2$ and $L_t=5$ according to the
 one-loop formulas.
 Therefore we neglected the uncertainties of 
 ${\partial \lambda}/{\partial \tau}$ and 
 ${\partial \beta}/{\partial \tau}$. 
 Together with the results for $\partial \kappa/\partial \tau$ in
 eq.~(\ref{eq14}), we obtain 
\begin{equation}\label{eq17}
\left.\frac{\Delta\epsilon}{T_c^4}\right|_{\rm m2} = 0.240(30+4)
\ ,\qquad
\left.\frac{\Delta\epsilon}{T_c^4}\right|_{\rm m4} = 0.28(3+9)
\ .
\end{equation}
 The result with index m4 (temporal lattice extension $L_t=4$) was
 obtained in the point with hopping parameter $\kappa_c=0.128235$
 corresponding to table~\ref{tab06}.
 (In table~\ref{tab01} this point has the index m4[24/35].)
 The first number in parenthesis is the error originating from the
 uncertainties of $\partial \kappa / \partial \tau$ and the
 statistical errors quoted in table~\ref{tab07}.
 The second one gives the influence of the error of the critical
 hopping parameter.
 In the case of $L_t=4$ a more precise estimate of the critical
 hopping parameter is obviously desirable, in order to get a smaller
 error of the latent heat.
 Within the large errors, the result of the $L_t=4$ lattice agrees
 with that of $L_t=2$, and hence with scaling.

                                                                                
\section{Relation between energy density and pressure} \label{sec4}
 The thermodynamical quantity $\delta \equiv \epsilon/3-P$, where
 $\epsilon$ is the energy density and $P$ the pressure, has been used
 in section~\ref{sec33} for the determination of the latent heat.
 It is an interesting quantity on its own, because it reflects the
 deviation of the thermodynamical equations of state from those of
 a free relativistic massless gas (photon gas).
 In the latter case $\delta=0$, thus $\epsilon/3=P$.
 The reason why it is advantageous to extract the latent heat from
 the jump of $\delta$ at the phase transition is its relatively
 simple expression in terms of global averages on the lattice:
\be \label{eq18}                                                                
\delta = \frac{1}{3} (TL_t)^4                                                   
\left\langle \frac{\partial\kappa}{\partial\tau} 
\cdot 8L_{\varphi,x\mu}               
- \frac{\partial\lambda}{\partial\tau} \cdot Q_x                                
- \frac{\partial\beta}{\partial\tau} \cdot 6P_{pl} \right\rangle \ .
\ee                                                                             
 Here $T$ is the temperature corresponding to the temporal lattice
 extension $L_t$.
 The quantities $L_{\varphi,x\mu},\; Q_x,\; P_{pl}$ are defined in
 eqs.~(\ref{eq03}) and (\ref{eq16}) and $\tau \equiv -\log(M_W)$
 is the lattice scale parameter.
 The derivatives as ${\partial\kappa}/{\partial\tau}$ etc.\ are
 taken along the line of constant physics going through the point
 $(\beta,\lambda,\kappa)$ in bare parameter space.
 The expression in eq.~(\ref{eq18}) still contains a divergent
 vacuum contribution, which has to be subtracted from the right-hand
 side.
 In numerical simulations the most convenient way is to subtract from
 the right-hand side the same expression on a lattice with large
 temporal extension corresponding to zero temperature.
 In this way one ends up with $\delta(T)/T^4$, because
 $\lim_{T \searrow 0} \delta(T)/T^4 =0$.

 Of course, it would be interesting to know $\delta$ as a function of
 the temperature.
 For instance, much above the phase transition temperature $T \gg T_c$
 it can be expected, at least for small quartic coupling $\lambda$,
 that masses can be neglected and interactions are weak, which
 corresponds to $\delta(T)/T^4 \ll 1$.
 (Note that $\delta(T)/T^4$ is dimensionless.)
 Above the phase transition temperature the system is in the symmetric
 phase, where perturbation theory cannot be applied; therefore one has
 to rely on non-perturbative methods.
 In the Higgs phase, well below $T_c$, resummed perturbation theory
 should give a good description.

 In a numerical simulation the simplest way to change the temperature
 is to change the temporal lattice extension $L_t$, which is
 proportional to the inverse temperature.
 Another possibility would be to change the lattice spacing in time
 direction for fixed $L_t$ and spatial lattice spacing, but this
 would require the introduction of different gauge couplings and
 hopping parameters in the timelike and spacelike directions.
 We choose to change $L_t$ and keep the parameters fixed.
 Therefore, we repeated the numerical simulations at the critical
 values of the hopping parameter for $L_t=2$ and $L_t=4$, with
 different temporal lattice extensions and unchanged $\beta,\lambda$.
 Since we have $L_t \geq 2$, this means that at the $L_t=2$
 point the possible temperatures are $T=2T_c/n$ with $n \geq 2$.
 At the $L_t=4$ point we also have two temperatures above $T_c$
 because the possible values are $T=4T_c/n$.
 The numerical simulation results for $\delta(T)/T^4$ are shown in
 figure~\ref{delta_t4}, where vertical error bars refer to the
 statistical errors from the averages, horizontal ones to the errors
 due to the uncertainty in the critical hopping parameter, and the
 shaded areas indicate the error coming from the derivatives
 ${\partial\kappa}/{\partial\tau}$.
 The latter errors result in overall vertical shifts of the curves.
 (More precisely, the effect of changing
 ${\partial\kappa}/{\partial\tau}$ is to a good approximation an
 overall multiplication, therefore differences as
 $\log(\delta(T_1)/T_1^4)-\log(\delta(T_2)/T_2^4)$ are better
 determined than the individual values.)
 As can be seen, the errors are substantial, especially in the
 Higgs phase below $T_c$.
 In the symmetric phase, where the errors are relatively small,
 the dominant feature is a rather fast decrease: at $T=2T_c$
 the quantity $\delta(T)/T^4$ is very small, which implies the
 relation $\epsilon/3 \simeq P$.

 It is also interesting to try to extract the derivative
 $\partial (\delta(T)/T^4)/\partial T$ by changing the hopping
 parameter $\kappa$ in a small range around $\kappa_c$ and
 keeping the other two bare parameters $\beta$ and $\lambda$ fixed.
 Previous experience \cite{PHYSLETT,NUCLPHYS} and comparison of the
 present results at points with shifted $\kappa$-values, such as
 m2[12/65] with m2[12/70] and m4[24/35] with m4[24/85], show that
 the renormalized parameters $R_{HW}$ and $g_R^2(M^{-1})$ change
 very little.
 The only substantial change is in the lattice spacing, which
 can be seen e.~g.\ on the change of the W-mass.
 We neglect the small changes in $R_{HW}$ and $g_R^2(M^{-1})$ and
 make the reasonable assumption that the LCP's are parallel in a
 small $\kappa$ interval.
 (That is, the partial derivatives ${\partial\kappa}/{\partial\tau}$
 etc. appearing in eq.\ (\ref{eq18}) are unchanged.)
 Thus the derivative $\partial (\delta(T)/T^4)/\partial T$ can be
 obtained from the derivatives of the global averages 
 $\langle L_{\varphi,x\mu} \rangle$, $\langle Q_x \rangle$ and
 $\langle P_{pl} \rangle$.
 The derivatives of these quantities with respect to $\kappa$
 can be determined in numerical simulations by the reweighting
 technique.
 Then the derivative with respect to $T$ is given by
\be \label{eq19}
\frac{\partial}{\partial T_r} 
\left.\frac{\delta(T)}{T^4}\right|_{\kappa=\kappa_c} \simeq
-\frac{1}{T_r} \left[\frac{da}{d\kappa}\right]^{-1}
\frac{\partial}{\partial\kappa} \frac{\delta(T)}{T^4} \ ,  
\ee
 where we set the lattice spacing at $\kappa_c$ to unity and 
 $T_r \equiv T/T_c$.
 The numerical results from comparing the W-mass at the points
 m2[12/65] with m2[12/70] and m4[24/35] with m4[24/85],
 respectively, are 
\be \label{eq20}
\left[\frac{da}{d\kappa}\right]^{-1}_{\rm m2}
= 0.00105(10) \ , \hspace{2em}
\left[\frac{da}{d\kappa}\right]^{-1}_{\rm m4}
=0.000209(17) \ .
\ee
 The obtained estimates of the derivative  
 $\partial (\delta(T)/T^4)/\partial T$ are illustrated in
 figure~\ref{delta_der}.
 Note that at the phase transition point the derivatives are
 determined in the corresponding metastable phases.

 The results of the numerical simulations on $\delta(T)$ will be
 compared with perturbation theory results in the next section.

\section{Comparison with the perturbative predictions} \label{sec5}

 In the previous sections we have presented a quantitative
 description of the electroweak phase transition for $M_H \simeq$ 34
 GeV on the lattice.
 By comparing data from lattice simulations for the SU(2)-Higgs 
 model with the perturbative results, one can hope to identify
 non-perturbative features and to achieve a better understanding of
 the electroweak phase transition.
 Therefore, in this section we shall present a comparison between data
 of the present lattice results and the perturbative predictions.

 We have determined the renormalized masses at zero temperature
 ($M_H$, $M_W$), critical temperature ($T_c$), jump in the order
 parameter ($\rho$), latent heat ($\Delta \epsilon$), and the relation
 between energy density and pressure.
 Besides these data, we also include the interface tension ($\sigma$)
 given by \cite{CSFOHEHE}.
 As usual, the dimensionful quantities are normalized by the proper
 power of the critical temperature.

 The most complete perturbative result for the four-dimensional finite
 temperature electroweak phase transition has been presented in 
 \cite{FOHE,BUFOHE}. 
 For increasing Higgs boson mass the perturbative prediction is
 less and less reliable, since the relative difference
 between the one-loop and the two-loop perturbative results grows.
 The deviation strongly depends on the observable: the critical
 temperature turns out to be quite insensitive to the
 loop order; however the interface tension receives corrections
 of ${\cal O}(100\%)$ even for $M_H \simeq$ 34 GeV.

 The present analysis follows the method of \cite{BUFOHE}. 
 As was mentioned in \cite{BUFOHE}, the treatment of the
 interface tension has only been performed in the resummation scheme
 of \cite{FOHE}, therefore this scheme will be applied here.
 In order to make this section self-contained, we recall our
 convention to treat the high temperature expansion and the
 renormalization scheme dependence.

 The $g^3,\lambda^{3/2}$-potential of \cite{FOHE} involves a
 high-temperature expansion up to order $(m/T)^3$, which is
 unsatisfactory for $M_H\simeq 34$ GeV.
 Thus, we have included all one-loop contributions of order $(m/T)^4$
 in our present $g^3,\lambda^{3/2}$-potential.
 Note, that the numerical evaluation of the one-loop temperature
 integrals gives a result which agrees with the above approximation
 up to a few percent.

 For the present Higgs boson mass the renormalization scheme
 dependence is non-negligible.
 We shall use the scheme suggested by Arnold and Espinosa
 \cite{ARNESP}.
 It includes the most important zero-temperature renormalization
 effects and is very close to the on-shell renormalization
 scheme, which is used by the lattice determination of the masses.
 Note, that this scheme has been used previously for $T_c/M_H$ in the
 insert of fig. 15 of ref.~\cite{NUCLPHYS} and for
 $\rho/T_c,\ \Delta \epsilon/T_c^4, \sigma/T_c^3$ and $T_c/M_H$ 
 in ref.~\cite{BUFOHE}.
 In this scheme the correction to the $\overline{\mbox{MS}}$-potential,
 used for both the one- and the two-loop results, reads
\begin{equation} \label{eq21}
\delta V={\varphi^2 \over 2} \left( \delta\mu + {1 \over 2\beta^2}
\delta\lambda\right) + {\delta\lambda \over 4} \varphi^4,
\end{equation}
 where
\begin{equation} \label{eq22}
\delta \mu = {9g^4v^2 \over 256 \pi^2},\ \ \ \ \ \
\delta\lambda=-{9g^4\over 256\pi^2}\left(\ln\frac{M_W^2}{\bar{\mu}^2}
+{2\over3} \right).
\end{equation}
 Here $v$ is the zero-temperature vacuum expectation value, $M_W$ the
 W-boson mass at $T=0$, and the form of the potential at $T=0$ is
\begin{equation} \label{eq23}
V={\varphi^2 \over 2}\mu + {\lambda \over 4}\varphi^4.
\end{equation}
%
\renewcommand{\arraystretch}{1.5}\label{CC}
\begin{table}
\begin{center}
\parbox{15cm}{\caption{\label{tab08}\it
 Comparison of the normalized latent heat obtained by using
 the Clausius--Clapeyron equation and the direct lattice formula.
}}
\end{center}
\begin{center}
\begin{tabular}{|l||l|l|}
\hline
& $L_t=2$  & $L_t=4$  \\ \hline\hline
$\Delta \epsilon/T_c^4$ from eq.~(\protect\ref{eq24})  & 
0.281(19) & 0.31(12) \\ 
\hline
$\Delta \epsilon/T_c^4$ from eq.~(\protect\ref{eq15})  & 
0.240(34) & 0.28(12) \\ \hline
\end{tabular}
\end{center}
\end{table}

 It is instructive to compare the direct lattice result for the 
 latent heat with the prediction of the Clausius--Clapeyron equation, 
 obtained in perturbation theory \cite{BUFOHE}:
\begin{equation} \label{eq24}
\Delta\epsilon \simeq -\kappa M_H^2 \Delta\rho^2 \ .
\end{equation}
 (The factor $\kappa$, instead of the more usual $1/2$, is due to
 our normalization conventions.)
 The first line of table~\ref{tab08} shows this, using the lattice
 results for $\Delta\rho^2$ and $M_H^2$.
 The second line contains the latent heat determined in
 section~\ref{sec33}.
 As usual, the numbers in parenthesis denote the errors.
 The values agree with the prediction of the Clausius--Clapeyron
 equation within one standard deviation.

 Let us compare the lattice results on the jump of the order parameter
 ($\rho$), latent heat, interface tension and critical temperature
 with the perturbative predictions (see fig.~\ref{perturb}). 
 For each quantity the dashed lines show the region allowed by the
 statistical error of a given lattice observable, whereas the dotted
 lines include an estimate of the  systematic error as well.
 Since the results with largest $L_t$ are closest to the continuum
 limit, we have plotted them.
 The statistical errors of the above lattice observables were
 determined by standard methods.
 In cases where the $L_t=2$ result significantly differs from those at
 larger $L_t$, rough estimates of the systematic errors can be
 obtained from these deviations.
 For the interface tension, where only $L_t=2$ data exist, the
 systematic error was estimated from the difference between the
 results of the ``transfer matrix'' and the ``two-couplings'' 
 methods \cite{CSFOHEHE}.
 The last quantity, namely $T_c/M_H$, will be discussed below.

 A rough estimate of the uncertainties of two-loop resummed 
 perturbation theory is given by the difference between the one-loop 
 (shown by triangles on the plot) and two-loop result (shown by
 squares). 
 In order to represent the region given by this uncertainty we have 
 connected the point of the one-loop result with that of the
 two-loop result.
 Comparing to the numerical results, an additional uncertainty
 arises since neither the Higgs boson mass nor the gauge coupling
 have been determined exactly.
 Therefore, the perturbative prediction for an observable at a given
 order is not one definite value but rather an interval, given by the
 uncertainties of $M_H$ and $g_R$. 
 These errors are small. 
 They are represented on the plot by error bars left and right to the
 triangles and squares, respectively.
 As a perturbative reference point we will use the values 
 $R_{HW}=0.422(11)$ and $g^2=0.585(10)$. 
 We neglect corrections due to the different renormalization 
 conditions used for coupling $g_R$ and the order parameter 
 $\rho$ on the lattice and in the continuum.
 These corrections are expected to be of relative order $g_R^2$.

 The inspection of figure~\ref{perturb} shows that for the jump of
 the order parameter, the latent heat, and the interface tension the
 agreement between numerical simulations and perturbation theory is
 good. 
 The errors are, however, not small, except for the simulation result
 for the jump of the order parameter and the perturbative prediction
 for the latent heat.
 The interface tension has huge corrections in perturbation theory;
 nevertheless, the two-loop result agrees with the lattice data.
 Here one has to note that the perturbative calculation of the
 interface tension is on different footing, and is up to now less
 understood, than those of the other quantities.
 (See e.~g.\ ref.~\cite{BBFH}.)
 In case of the ratio of the transition temperature to the Higgs
 boson mass $T_c/M_H$, where the results of one-loop and two-loop
 perturbation theory almost coincide, the numerical simulation results
 and the extrapolation to the continuum limit are collected in table
 \ref{tab09} and shown in figure~\ref{extra}.
 The errors of the simulations for this quantity are dominated by
 the uncertainties in the critical hopping parameter discussed in
 section~\ref{sec3}.
 (This is in contrast to $R_{HW}$ and $g_R^2(M^{-1})$, where these
 errors cancel to a large extent.)
 The extrapolated continuum value is $T_c/M_H=2.147(40)$.
 The value at $L_t=2$ is about 5\% smaller.
 This relatively small deviation is better than the expectation
 based on lattice perturbation theory.
 For instance, the third paper of ref.~\cite{FAKARUSH} gives an
 estimate of scaling violations of about 20-30\% for $L_t=2$.
 As figure~\ref{extra} shows, the value of $T_c/M_H$ extrapolated to
 the continuum limit differs by about three standard deviations from
 the two-loop perturbative result.
 This is under the assumption that the $L_t=2$ point can be included
 in the extrapolation, which is supported by the good quality of the
 fit ($\chi^2 \simeq 1$) and the smallness of scaling violations
 also discussed in section \ref{sec2}.
 In our opinion, one cannot exclude the possibility that there are
 some non-negligible higher-loop contributions and/or
 non-perturbative effects.
 These could show up also in other quantities, once the errors
 there get similarly small.
\begin{table}
\begin{center}
\parbox{15cm}{\caption{\label{tab09}\it
 Relation of the critical temperatur $T_c$ and the masses $M_H$ and
 $M_W$.
 For the continuum limit lattice artefacts of order
 ${\mathcal O}(a^2)$ were considered.
}}
\end{center}
\begin{center}\begin{tabular}{|c||llll|l|}
\hline 
$a$ & \multicolumn{1}{c}{$[2T_c]^{-1}$}&
 \multicolumn{1}{c}{$[3T_c]^{-1}$}
      & \multicolumn{1}{c}{$[4T_c]^{-1}$}&
 \multicolumn{1}{c|}{$[5T_c]^{-1}$}& 
            \multicolumn{1}{c|}{${a\to 0}$}\\ 
\hline \hline
$T_c/M_H$& 2.026(11) & 2.100(33) & 2.087(48) & 2.21(10)& 2.147(40)\\ 
\hline
$T_c/M_W$& 0.8844(30)& 0.898(13) & 0.904(19) & 0.905(40)& 0.910(16)\\ 
\hline
\end{tabular}\end{center}
\end{table}


 An interesting new possibility to compare perturbation theory and
 numerical simulations {\em as a function of the temperature} is to
 consider $\delta(T)/T^4$ investigated in the previous section.
 Since the perturbative calculations also involve a high temperature
 expansion, in this framework it is not possible to perform a
 subtraction at $T=0$.
 In the symmetric phase, due to the unsolved infrared problems,
 perturbation theory is not applicable.
 In figure~\ref{delta_pert} the comparison is done for
 $(\delta(T)-\delta(T_c))/T_c^4$, where $\delta(T_c)$ is taken in
 the Higgs phase.
 The agreement is reasonable but not perfect.
 In fact, one can question the reliability of perturbation theory
 already at $T_c$, therefore a subtraction at, say, $T=T_c/2$
 would be even safer.
 In this case the curve in figure~\ref{delta_pert} is shifted downwards
 and inside the Higgs phase the agreement becomes good.
 The discrepacy is then shifted to the point at $T_c$.

                                                                                
\section{Discussion}                                   \label{sec6}
 The results presented in this paper are the outcome of detailed
 numerical simulations of the thermodynamical properties of the 
 electroweak phase transition in the SU(2) Higgs model at Higgs boson
 mass $M_H \simeq 34$ GeV.

 The high precision data for the correlation functions and Wilson
 loops were fitted carefully to extract the true statistical
 errors of the masses and static potential, respectively.
 We also tried to identify and control the systematic errors such as
 finite volume effects and lattice artefacts.
 The former are well under control and we could extrapolate the
 results to infinite volume reliably.
 In order to estimate lattice artefacts, we performed the first
 high statistics numerical simulations with temporal lattice
 extensions up to $L_t=5$.
 As the results show, for interesting physical quantities the
 deviations between $L_t=2$ and the maximal investigated $L_t$ are
 small (see eq.~\ref{eq17}, table \ref{tab08}, fig.~\ref{extra}).
 In general, the size of lattice artefacts turned out to be
 surprisingly small, in accordance with previous observations
 concerning $L_t=2$ and $L_t=3$ \cite{PHYSLETT,NUCLPHYS}.
 Therefore, once sufficiently small statistical errors are achieved,
 the extrapolation to the continuum limit from the range
 $2 \leq L_t \leq 5$ seems feasible.
 Note in this respect that our largest errors in some important cases
 come from the uncertainties in the critical hopping parameters,
 which are due to the use of 32-bit arithmetics in the simulations.

 It is interesting to compare the numerical simulation results to
 perturbation theory, which is expected to work well at
 $M_H \simeq 34$ GeV for some quantities ($T_c$ or $\rho$); however,
 this Higgs boson mass appears to be at the edge of its domain
 of validity for other quantities (e.g. $\Delta \epsilon$ or
 $\sigma$).
 As the discussion in section~\ref{sec5} shows, there is in general
 a satisfactory agreement within the present accuracy.
 However, an important discrepancy is observed in the simple quantity
 $T_c/M_H$ at the level of about 3 standard deviations, which could
 perhaps hint to higher-order and/or non-perturbative contributions
 (see figure~\ref{extra}).
 In fact this quantity is relatively easy to determine with small
 errors.
 It is possible that deviations of similar magnitude will emerge in
 other quantities as well, once the precision becomes similar there.

\begin{figure}
\epsfig{file=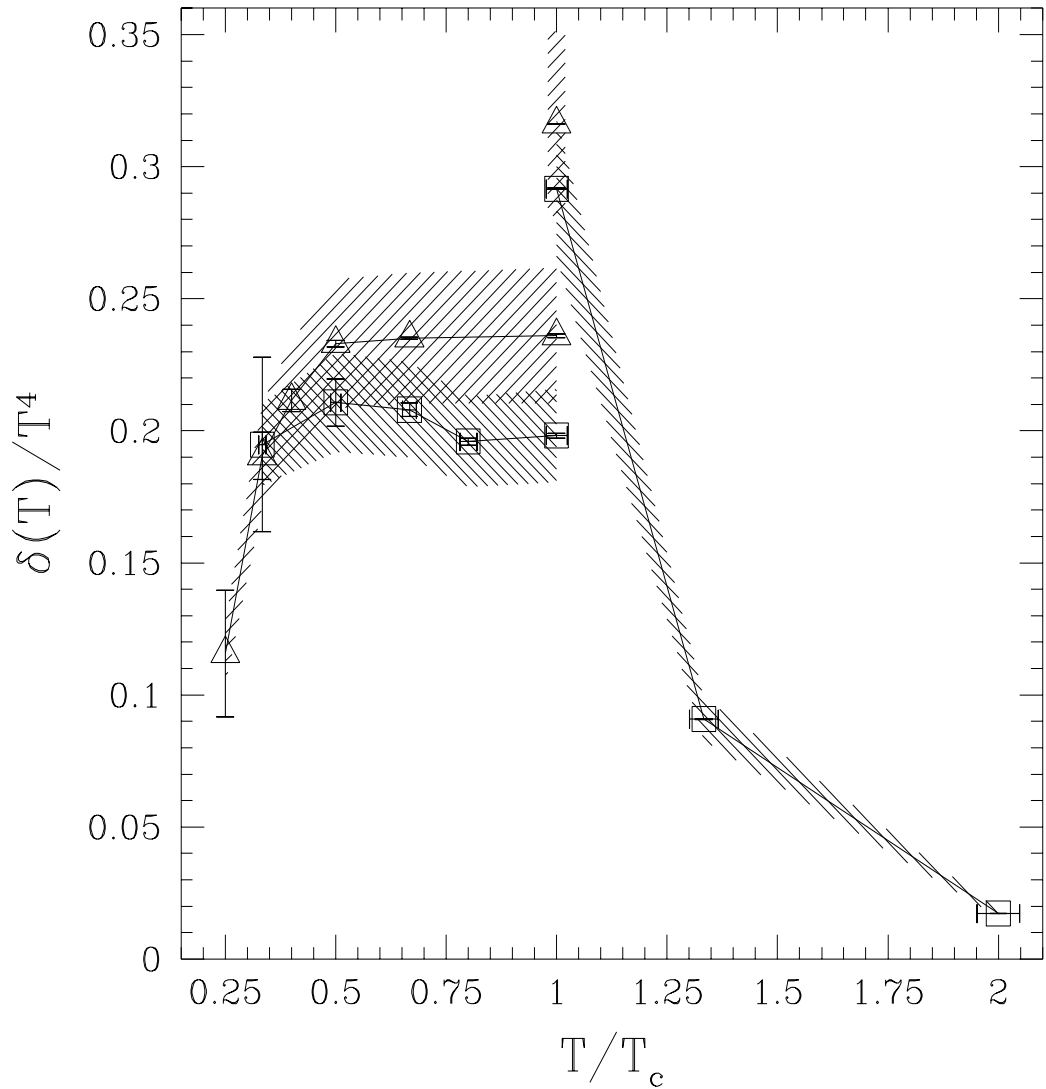,width=17.0cm,height=15.0cm,
        bbllx=0pt,bblly=200pt,bburx=600pt,bbury=700pt,angle=0}
\begin{center}
\parbox{15cm}{\caption{ \label{delta_t4}
 Results of the numerical simulations for $\delta(T)/T^4$ on $L_t=2$
 and $L_t=4$ lattices.
 The former are shown by triangles the latter by boxes.
 The errors are explained in the text.
}}
\end{center}
\end{figure}

\begin{figure}
\epsfig{file=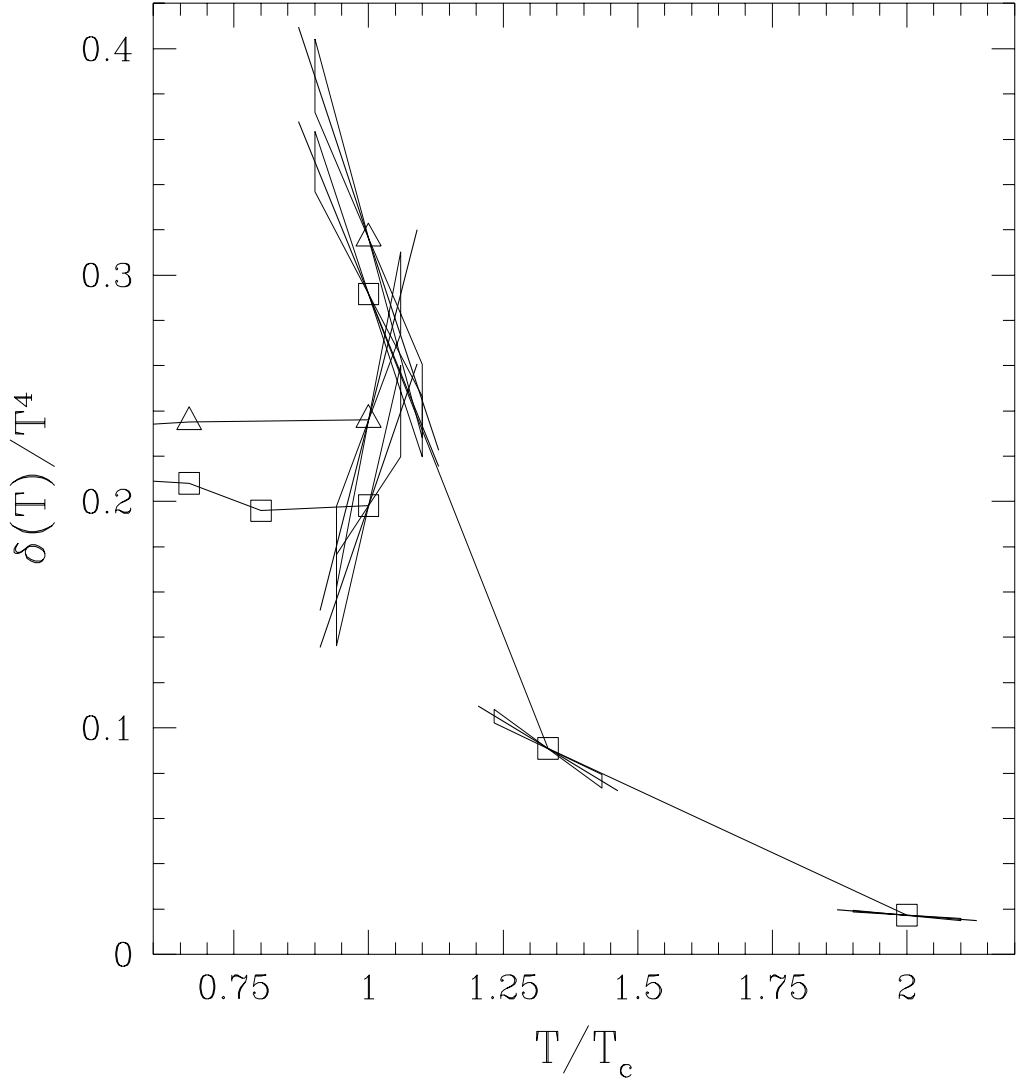,width=17.0cm,height=15.0cm,
        bbllx=0pt,bblly=200pt,bburx=600pt,bbury=700pt,angle=0}
\begin{center}
\parbox{15cm}{\caption{ \label{delta_der}
 $\delta (T)/T^4 $ as a function of $T/T_c $ together with 
 $\partial (\delta (T)/T^4 )/\partial T$. 
 The straight lines represent the tangents as calculated by the method
 described in the text (their lengths have no significance). 
 For $T/T_c < 1$ the results on the derivatives are consistent with
 zero, and are not shown.
 The elongated triangles around the straight lines indicate the errors
 of the tangents.}}
\end{center}
\end{figure}

\begin{figure}
\epsfig{file=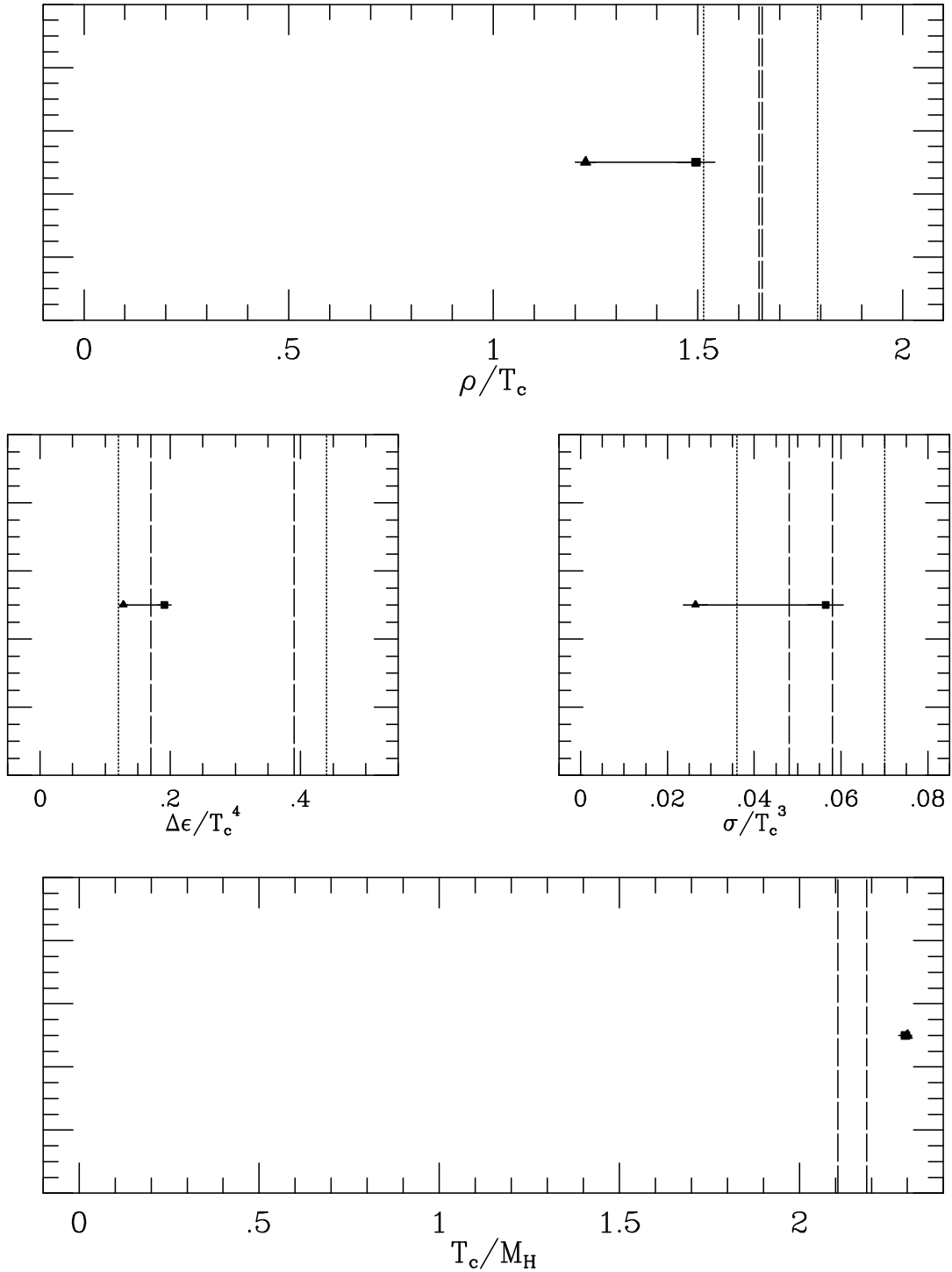,width=17.0cm,height=20.0cm,
        bbllx=0pt,bblly=50pt,bburx=600pt,bbury=750pt,angle=0}
\begin{center}
\parbox{15cm}{\caption{ \label{perturb}
 Comparison of the numerical simulation results with those from
 two-loop resummed perturbation theory.
 Vertical strips give the numerical results with errors.
 The triangles are the perturbative predictions at order $g^3$,
 the boxes those at $g^4$.
}}
\end{center}
\end{figure}

\begin{figure}
\epsfig{file=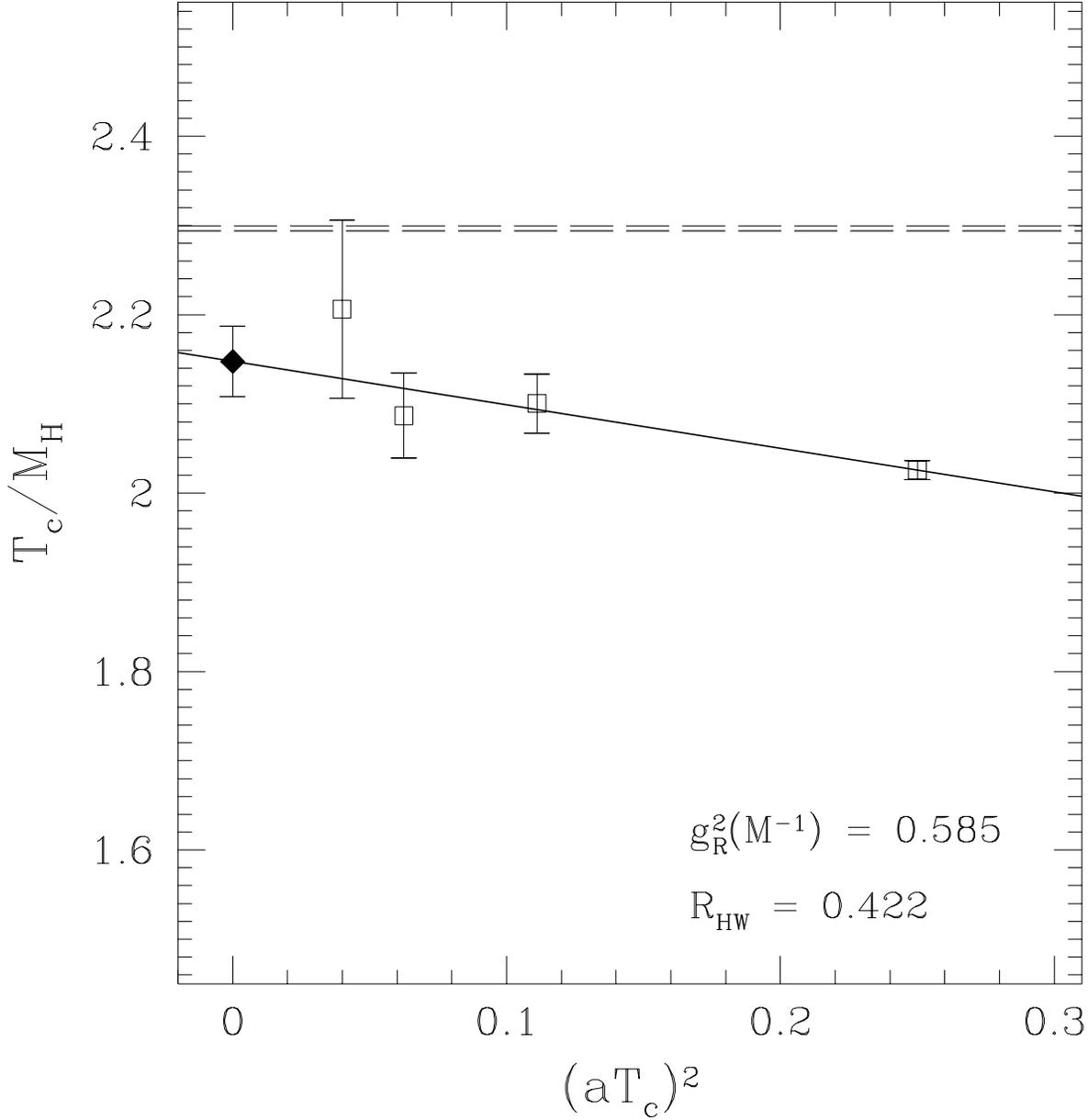,width=17.0cm,height=17.0cm,
        bbllx=75pt,bblly=175pt,bburx=600pt,bbury=700pt,angle=0}
                                           
\begin{center}
\parbox{15cm}{\caption{ \label{extra}
 The numerical results for the ratio of the transition temperature
 and the Higgs boson mass $T_c/M_H$ versus $(aT_c)^2=L_t^{-2}$.
 The straight line is the extrapolation to very small lattice
 spacings, which gives the continuum value shown by the filled
 symbol. 
 The dashed horizontal lines are the perturbative predictions at order
 $g^3$ (upper) and $g^4$ (lower), respectively. 
}}
\end{center}
\end{figure}

\begin{figure}
\epsfig{file=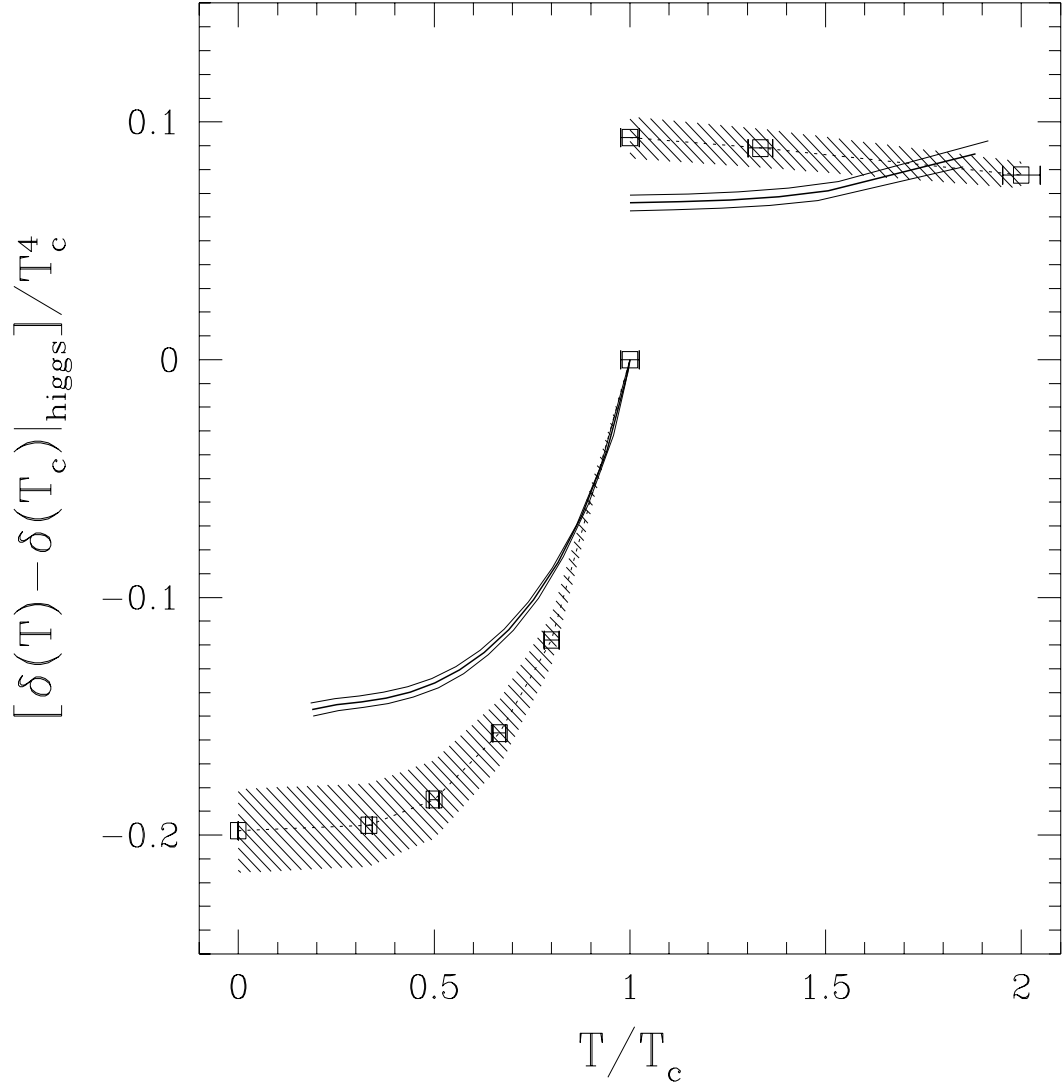,width=17.0cm,height=15.0cm,
        bbllx=0pt,bblly=200pt,bburx=600pt,bbury=700pt,angle=0}
\begin{center}
\parbox{15cm}{\caption{ \label{delta_pert}
 Comparison of the numerical results for $\delta(T)$ with those from 
 two-loop perturbation theory.
 The shaded areas show the uncertainty of the numerical simulation
 results due to the uncertainty in the derivatives of bare parameters
 along the lines of constant physics.
 The solid lines show the perturbative results together with the
 uncertainties induced by the errors on $R_{HW}$ and $g_R^2$.
}}
\end{center}
\end{figure}

 We also determined the temperature dependence of the thermodynamical
 quantity $\delta \equiv \epsilon/3-P$ ($\epsilon$ = energy density,
 $P$ = pressure) in a range $1/4 \leq T/T_c \leq 2$ (see
 figure~\ref{delta_t4}).
 In the Higgs phase the errors are relatively large.
 Within these errors there is a reasonable agreement with perturbation
 theory.
 In the symmetric phase, where perturbation theory is plagued by
 uncontrollable infrared singularities, $\delta/T^4$ becomes rapidly
 rather small: at the last point $T/T_c=2$ it is already almost
 compatible with zero.
 Therefore, at higher temperatures we have $\epsilon/3 \simeq P$, as
 in the photon gas.
 The knowledge of this thermodynamical equation of state is important
 in the history of the early Universe.

\vspace{10mm}                                                    
{\large\bf Acknowledgements} 

\vspace{5mm}\noindent
 We thank W. Buchm\"uller, A. Hebecker, K. Jansen, A. Patk\'os and
 R. Sommer for discussions.
 Two of us (F.~Cs. and Z.~F.) were partially supported by Hungarian 
 Science Foundation grant under Contract No.\ OTKA-F1041/3-T016248/7.


\newpage
                                                                                
\end{document}